\documentclass[preprint, dvipdfmx]{aastex}
\slugcomment{accepted for publication in ApJ}
\shorttitle{A comparison of SPH AVs and their impact on the Keplerian disk}
\shortauthors{Hosono et al.}

\renewcommand{\vec}[1]{\mbox{\boldmath $#1$}}
\newcommand{\mat}[1]{\mathbf{\mathsf{#1}}}

\newcommand{\step}[2]{\textbf{Step#1}:}

\begin{document}
\title{A comparison of SPH artificial viscosities and their impact on the Keplerian disk}

\author{\textsc{Natsuki Hosono}\altaffilmark{1,*}, \textsc{Takayuki R. Saitoh}\altaffilmark{2}, \textsc{Junichiro Makino}\altaffilmark{1,2}}
\altaffiltext{1}{RIKEN Advanced Institute for Computational Science, Minatojima-minamimachi, Chuo-ku, Kobe, Hyogo 650-0047, Japan}
\altaffiltext{2}{Earth-Life Science Institute, Tokyo Institute of Technology, Ookayama, Meguro-ku, Tokyo 152-8550, Japan}
\altaffiltext{*}{Tel: +81-(0)78-940-5707 / Fax: +81-(0)78-304-4972}
\email{natsuki.hosono@riken.jp}

\begin{abstract}
Hydrodynamical simulations of rotating disk play important roles in the field of astrophysical and planetary science.
Smoothed Particle Hydrodynamics (SPH) has been widely used for such simulations.
It, however, has been known that with SPH, a cold and thin Kepler disk breaks up due to the unwanted angular momentum transfer.
Two possible reasons have been suggested for this breaking up of the disk; the artificial viscosity (AV) and the numerical error in the evaluation of pressure gradient in SPH.
Which one is dominant has been still unclear.
In this paper, we investigate the reason for this rapid breaking up of the disk.
We implemented most of popular formulations of AV and switches and measured the angular momentum transfer due to both AV and the error of SPH estimate of pressure gradient.
We found that the angular momentum transfer due to AV at the inner edge triggers the breaking up of the disk.
We also found that the classical von-Neumann-Richtmyer-Landshoff type AV with a high order estimate for $\vec{\nabla} \vec{\cdot} \vec{v}$ can maintain the disk for $\sim 100$ orbits even when used with the standard formulation of SPH.
\end{abstract}


\section{Introduction}
The hydrodynamical simulations of rotating disk play important roles in the field of astrophysical and planetary science, since astronomical objects are formed by the gravitational collapse and the conservation of initially imprinted tiny angular momentum makes rotating disks during the collapsing process, e.g., galactic disks, accretion disks, protoplanetary disks and post impact debris disks.
Smoothed Particle Hydrodynamics \citep[SPH][]{L77, GM77, M92, S10} method has been widely used for these simulations.

However, it has been long known that SPH cannot follow the long-term evolution of thin and cold disks \citep[e.g.,][]{M+96, II02, O+03}.
There are two possible reasons for the decay of the disk.
One is the error in the discretized hydrodynamical force (pressure gradient), which is often referred to as the ``E${}_0$ error'' or ``zero-th order error'' in the SPH discretization \citep[e.g.,][]{D99, II02, O+03, R+10}.
\citet{O+03} pointed out that the hydrodynamical torque leads to the disruption of the galactic disk embedded in a hot halo.

The other possible reason is the artificial viscosity (AV).
AV is a numerical dissipation term which is required to capture the shock, first proposed by \citet{R48} and \citet{VR50}.
Since AV does not exist in the original governing equations, AV should act only on the shock.
In SPH, there are two well known formulations of AV.
One is based on the ``$\vec{\nabla} \vec{\cdot} \vec{v}$'' term where $\vec{v}$ means velocity, and the other is based on relative velocities of neighbor particles (see sections \ref{sec:vNRL_AV} and \ref{sec:pairwise_AV} for detail).
In the following, we call the former one ``$\vec{\nabla} \vec{\cdot} \vec{v}$'' AV and the latter one ``pairwise'' AV.
In almost all recent works, the latter ``pairwise'' AV has been used.
It has been pointed out that, however, AV, in particular the latter formulation, operates not only on the shock, but also on the velocity shears.
This spurious shear viscosity causes unwanted angular momentum transfer in rotating systems.
To suppress this effect, several ``switches'' to suppress AV in shear flow have been proposed \citep[e.g.,][]{B95, CD10}.
\citet{B+16}  have compared their implementation of SPH with some of these switches and the ``standard'' SPH in the context of galaxy formation.

Recently, \citet{GN11} and \citet{H15} developed a new scheme, Weighted Particle Hydrodynamics (WPH).
Unlike SPH, WPH adopts the Riemann solver to deal with the shock, which introduces sufficient viscosity at the shock.
In addition, they argued that WPH resolved the zero-th order error.
\citet{H15} reported that WPH could maintain the thin and cold rotational disk for much longer time than SPH can.

Despite these efforts, the main reason for the rapid disruption of the disk is still unclear.
In this paper, to clarify the reason of the disruption of the disk, we compare most of popular formulations of AV, their switches and derivative operators.
We performed the systematic survey of all possible combinations of different formulations.
We then investigate the angular momentum transfer due to both AV and SPH errors.

This paper is organized as follows.
In Sec. 2, we overview the idea of AV, its numerical formulations and switches.
In Sec. 3, we show the results of the comparison test.
In Sec. 4, we summarize the reasons for the disruption of the disk.

\section{The artificial viscosity}
In this section, we describe the formulation and implementation of AV used in this paper.
The formula for the pressure gradient evaluation used in SPH method are given in Appendix \ref{Sec:App0}.

In this paper we discuss two well-known formulations of AV.
In section \ref{sec:vNRL_AV}, we present the classical von-Neumann-Richtmyer-Landshoff (vNRL) AV which is based on the discretized estimate of $\vec{\nabla} \vec{\cdot} \vec{v}$ \citep{R48, VR50, L55, L77}.
In section \ref{sec:pairwise_AV}, we present the more widely used form of AV, based on the pairwise relative velocity of particles \citep{MG83, M97}.
In sections \ref{sec:BulkSwitch} and \ref{sec:ShearSwitch}, we discuss the schemes proposed to reduce the strength of AV in regions without shocks.
In section \ref{sec:BulkSwitch}, we discuss time-dependent AVs \citep{MM97, CD10, RH12, R15}, and in section \ref{sec:ShearSwitch}, so-called shear switches \citep{B95, CD10}.

\subsection{von-Neumann-Richtmyer-Landshoff AV to SPH}\label{sec:vNRL_AV}
\subsubsection{formulation}
In the vNRL AV term, artificial ``pressure'' term, $p^\mathrm{AV}$ is added to the pressure in SPH equations (Eqs. \ref{eq:motion} and \ref{eq:energy}).
The artificial pressure $p^\mathrm{AV}$ is given by
\begin{eqnarray}
p^\mathrm{AV}_i = \left\{ \begin{array}{ll}
	- \alpha^\mathrm{AV}_i \rho_i c_i h_i (\vec{\nabla} \vec{\cdot} \vec{v})_i + \beta^\mathrm{AV}_i \rho_i h_i^2 (\vec{\nabla} \vec{\cdot} \vec{v})_i^2 & \mathrm{if} \, \left( \vec{\nabla} \vec{\cdot} \vec{v}\right)_i < 0, \\
	0 & \mathrm{otherwise},
\end{array} \right. \label{eq:local_AV}
\end{eqnarray}
where $\rho, c$ and $h$ are the density, the sound speed and the smoothing length, respectively.
The parameters $\alpha^\mathrm{AV}$ and $\beta^\mathrm{AV}$ determine the strength of the viscosity.
Typically $\alpha^\mathrm{AV} = 1$ and $\beta^\mathrm{AV} = 2 \alpha^\mathrm{AV}$ are used.
This approach have been tested by \citet{MG83} and \citet{HK89}.

The corresponding timestep $\Delta t$ for this AV is given by \citep{HK89}:
\begin{eqnarray}
\Delta t_i^\mathrm{CFL} = C^\mathrm{CFL} \frac{h_i}{h |\vec{\nabla} \vec{\cdot} \vec{v}_i| + c_i + 1.2 (\alpha_i c_i + \beta_i h_i |\min(\vec{\nabla} \vec{\cdot} \vec{v}_i, 0)|)},
\end{eqnarray}
where $C^\mathrm{CFL}$ is a CFL coefficient which is set to $0.3$ in this paper.

\subsubsection{The discretization of $\vec{\nabla} \vec{\cdot} \vec{v}$ term}\label{sec:DerivativeOperator}
In this section we discuss the methods used to calculate the $\vec{\nabla} \vec{\cdot} \vec{v}$ term.
To evaluate $\vec{\nabla} \vec{\cdot} \vec{v}$, the following scheme is widely used \citep[e.g.,][]{M92}:
\begin{eqnarray}
\vec{\nabla} \vec{\cdot} \vec{v}_i = \frac{1}{\rho_i} \sum_j m_j (\vec{v}_j - \vec{v}_i) \vec{\cdot} \vec{\nabla} W(\vec{x}_j - \vec{x}_i; h_i), \label{eq:SPH_div}\
\end{eqnarray}
where $m$ and $\vec{x}$ are the mass and position vector.

Recently, \citet{G+12} proposed a discretization of derivative operators more accurate than those used in the standard SPH discretization.
In \citet{G+12}, however, they did not applied their derivative to vector fields.
Here we extend their derivative estimate to $\vec{\nabla} \vec{\cdot} \vec{v}$ as
\begin{eqnarray}
\vec{\nabla} \vec{\cdot} \vec{v}_i & = & \sum_j \frac{m_j}{\rho_j} (\vec{v}_j - \vec{v}_i) \vec{\cdot} \left[ \mat{M}^{-1}_i (\vec{x}_j - \vec{x}_i) \right] W(\vec{x}_j - \vec{x}_i; h_i), \label{eq:G+12_div}\\
\mat{M}_i & = & \sum_j \frac{m_j}{\rho_j} (\vec{x}_j - \vec{x}_i) \vec{\otimes} (\vec{x}_j - \vec{x}_i) W(\vec{x}_j - \vec{x}_i; h_i).
\end{eqnarray}
The derivation of Eq. (\ref{eq:G+12_div}) is given in Appendix \ref{Sec:App1}.

Note that \citet{P05} also derived a derivative operator similar to Eq. (\ref{eq:G+12_div}).
We have also tested the \citet{P05}'s derivative operator.
However, the results were similar to those obtained with Eq. (\ref{eq:G+12_div}).
Thus, we do not show the results with the \citet{P05}'s derivative operator.

Note that we also need $\vec{\nabla} \vec{\times} \vec{v}$ and/or $\vec{\nabla} \vec{\otimes} \vec{v}$ in some cases (see, section 2.3 and 2.4).
By replacing the operator $\vec{\cdot}$ in Eqs. (\ref{eq:SPH_div}) and (\ref{eq:G+12_div}) with $\vec{\times}$ or $\vec{\otimes}$, we can easily obtain the discretized expression for $\vec{\nabla} \vec{\times} \vec{v}$ and $\vec{\nabla} \vec{\otimes} \vec{v}$.

\subsection{The pairwise artificial viscosity}\label{sec:pairwise_AV}
The ``pairwise'' formulation of AV \citep{MG83, M97} is based on the relative velocities between neighboring particles.
This AV has been the most widely used form of AVs in SPH, e.g., \texttt{GASOLINE} \citep{W+04}, \texttt{Gadget2} \citep{S05}, \texttt{MAGMA} \citep{RP07}, \texttt{VINE} \citep{W+09, N+09}, \texttt{EvoL} \citep{M+10} and \texttt{SEREN} \citep{H+11}.
\citet{M97} derived this pairwise form of AV from the analogy to the Riemann Solver.
In this paper, we use the \citet{M97}'s AV as the representative for the pairwise AV:
\begin{eqnarray}
\Pi_{ij} & = & \left\{ \begin{array}{ll}
	- \displaystyle\frac{\alpha^\mathrm{AV}_{ij}}{2} \frac{v_{ij}^\mathrm{sig} w_{ij}}{\rho_{ij}} \quad (w_{ij} < 0), \\
	0 \quad (\mathrm{otherwise}),
\end{array} \right. \label{eq:pairwise_AV}\\
\alpha_{ij}^\mathrm{AV} & = & \frac{\alpha^\mathrm{AV}_i + \alpha^\mathrm{AV}_j}{2},\\
w_{ij} & = & \frac{ (\vec{r}_j - \vec{r}_i) \vec{\cdot} (\vec{v}_j - \vec{v}_i) }{|\vec{r}_j - \vec{r}_i|},\\
v_{ij}^\mathrm{sig} & = & c_i + c_j - 3 w_{ij},\\
\rho_{ij} & = & \frac{\rho_i + \rho_j}{2},
\end{eqnarray}
where $\Pi_{ij}$ is the pairwise viscosity term.
The acceleration and heating due to this AV term is expressed as follows:
\begin{eqnarray}
\vec{a}_i^\mathrm{AV} & = & - \sum_j m_j \Pi_{ij} \frac{1}{2} \left(\frac{\vec{\nabla} W(\vec{x}_i - \vec{x}_j; h_i)}{\Omega_i} + \frac{\vec{\nabla} W(\vec{x}_i - \vec{x}_j; h_j)}{\Omega_j}\right),\\
\dot{u}_i^\mathrm{AV} & = & \frac{1}{2} \sum_j m_j \Pi_{ij} (\vec{v}_{i} - \vec{v}_j) \vec{\cdot} \frac{\vec{\nabla} W(\vec{x}_i - \vec{x}_j; h_i)}{\Omega_i},
\end{eqnarray}
where $\vec{a}, u$ and $p$ are the acceleration, specific internal energy and pressure for the particle $i$.
The function $W$ is the kernel function and $h$ is the smoothing length, which determine the spread of a particle and $\Omega$ is the so-called ``grad-h'' term \citep{SH02, H13}.
The timestep $\Delta t$ corresponding to this AV is \citep{M97}:
\begin{eqnarray}
\Delta t_i^\mathrm{CFL} = C^\mathrm{CFL} \frac{2 h_i}{\max_j v^\mathrm{sig}_{ij}}.
\end{eqnarray}

In addition to the timestep determined from AV, we also need to incorporate the timestep determined from the acceleration itself;
\begin{eqnarray}
\Delta t_i^\mathrm{Acc} = C^\mathrm{Acc} \sqrt{ \frac{h_i}{|\vec{a}_i|} }.
\end{eqnarray}
We set $C^\mathrm{Acc} = 0.3$.
Then, the actual timestep for particle $i$ is given by
\begin{eqnarray}
\Delta t_i = \min(\Delta t_i^\mathrm{CFL}, \Delta t_i^\mathrm{Acc}).
\end{eqnarray}

\subsection{Shock indicator}\label{sec:BulkSwitch}
\citet{MM97} proposed to vary $\alpha^\mathrm{AV}$ so that AV works only on the shock.
The basic idea of this approach is to increase $\alpha^\mathrm{AV}$ when the shock is approaching, and gradually reduce $\alpha^\mathrm{AV}$ otherwise.
\citet{R+00} suggested the following time derivative for $\alpha^\mathrm{AV}$:
\begin{eqnarray}
\frac{d \alpha^\mathrm{AV}}{dt} & = & \left( \alpha_{\max}^\mathrm{AV} - \alpha^\mathrm{AV} \right) \max \left(- \vec{\nabla} \vec{\cdot} \vec{v}, 0 \right) - \frac{\alpha^\mathrm{AV} - \alpha_{\min}^\mathrm{AV}}{\tau}, \label{eq:MM97}\\
\tau & \sim & \frac{h}{c}.
\end{eqnarray}
If $\vec{\nabla} \vec{\cdot} \vec{v}$ is negative, $\alpha^\mathrm{AV}$ is increased with the timescale inversely proportional to $\vec{\nabla} \vec{\cdot} \vec{v}$, up to the maximum value of $\alpha_{\max}^\mathrm{AV}$.
In this paper, we set $\alpha_{\max}^\mathrm{AV} = 2$ and $\alpha_{\min}^\mathrm{AV} = 0.1$.

\citet{CD10} suggested a higher order shock indicator.
In their approach, the time derivative of $\vec{\nabla} \vec{\cdot} \vec{v}$ is used as the shock indicator;
\begin{eqnarray}
\alpha_i^\mathrm{AV} & = & \max \left[ \alpha_{\max}^\mathrm{AV} \frac{A_i}{A_i + \max_j(v_{ij}^\mathrm{sig})^2/h_i^2}, 0 \right], \label{eq:CD10b_st}\\
A & = & \max \left[- f \frac{d (\vec{\nabla} \vec{\cdot} \vec{v})}{dt}, 0 \right], \\
\frac{d (\vec{\nabla} \vec{\cdot} \vec{v})}{dt} & = & \mathrm{tr} \left[ \vec{\nabla} \vec{\otimes} \vec{a} - (\vec{\nabla} \vec{\otimes} \vec{v})^2 \right], \\
\frac{d \alpha^\mathrm{AV}}{dt} & = & \frac{\alpha^\mathrm{AV} - \alpha_{\min}^\mathrm{AV}}{\tau},\label{eq:CD10b_to}
\end{eqnarray}
where $f$ is the shear switch described in the next section.

Other shock indicators have also been proposed.
\citet{RH12} used $\vec{\nabla} (\vec{\nabla} \vec{\cdot} \vec{v})$ for the shock indicator.
\citet{R15} combined both the \citet{CD10}'s and the \citet{RH12}'s approaches.
In this paper, for simplicity, we only show the results for \citet{R+00} and \citet{CD10} shock indicators.

\subsection{The shear switches}\label{sec:ShearSwitch}
The pairwise AV we discussed in section \ref{sec:pairwise_AV} has one critical drawback.
Since it operates whenever two neighboring particles are approaching, it works as shear viscosity.
In order to follow the evolution of differentially rotating disk, shear viscosity should be suppressed.
\citet{B95} proposed a switch to reduce AV when the divergence of the velocity is smaller than the rotation:
\begin{eqnarray}
f_i = \frac{|\vec{\nabla} \vec{\cdot} \vec{v}_i|}{|\vec{\nabla} \vec{\cdot} \vec{v}_i| + |\vec{\nabla} \vec{\times} \vec{v}_i| + \varepsilon c_i / h_i}, \label{eq:B95}
\end{eqnarray}
where $\varepsilon$ is a small value used to prevent the division by zero.
In this paper we set $\varepsilon = 10^{-4}$.
By adopting this switch, Eqs. (\ref{eq:local_AV}) and (\ref{eq:pairwise_AV}) can be rewritten as
\begin{eqnarray}
p_i^\mathrm{AV} & \longrightarrow & f_i p_i^\mathrm{AV},\\
\Pi_{ij} & \longrightarrow & \frac{f_i + f_j}{2} \Pi_{ij}.
\end{eqnarray}

\citet{CD10} suggested an alternative form of the shear switch:
\begin{eqnarray}
f_i & = & \frac{|2 (1 - R_i)^4 \vec{\nabla} \vec{\cdot} \vec{v}_i|^2}{|2 (1 - R_i)^4 \vec{\nabla} \vec{\cdot} \vec{v}_i|^2 + \mathrm{tr}(\mat{S}_i \mat{S}_i^\mathrm{t})}, \label{eq:CD10s_st}\\
R_i & = & \frac{1}{\rho_i} \sum_j \mathrm{sign} (\vec{\nabla} \vec{\cdot} \vec{v}_j) m_j W(\vec{x}_i - \vec{x}_j; h_i),\\
\mat{S} & = & \frac{1}{2} \left[ \vec{\nabla} \vec{\otimes} \vec{v} + (\vec{\nabla} \vec{\otimes} \vec{v})^\mathrm{t} \right] - \frac{1}{\nu} \left( \vec{\nabla} \vec{\cdot} \vec{v} \right) \mat{I}, \label{eq:CD10s_to}
\end{eqnarray}
where $\nu$ is the number of dimensions and $\mat{I}$ is the identifity matrix.

\section{Keplerian disk test}\label{Sec:test}
As shown in Table \ref{tab:table1}, we have two options for the form of AV, three for the shock indicator, three for the shear switch and two for the formula for the discretization of $\vec{\nabla} \vec{\cdot} \vec{v}$.
Therefore, we have $2 \times 3 \times 3 \times 2 = 36$ possible combinations of different schemes.
In this section, we show the results of 2D Keplerian tests for all these 36 AV implementations.
Each run is labeled as (vNRL,~M97)-(No,~R+00,~CD10)-(No,~B95,~CD10)-(SPH,~G+12).
The first, second, third and fourth options are AV form, shock indicator, shear switch and derivative operators, respectively.
Note that once we select the derivative operator, all $\vec{\nabla} \vec{\cdot} \vec{v}$ terms in the shock indicator and the shear switch are replaced with the selected discretized expression.
In this paper, we tested two SPH formulations, the standard SPH (SSPH) \citep[for review, see][]{M92} and Density Independent SPH (DISPH) \citep{SM13, H13}.
We, however, note that the results and conclusions are the same between two methods.
Thus, in the following we on only show the results with DISPH.

\begin{table}[htb]
	\caption{List of the implementations for AV tested in this study.}
	\begin{tabular}{r||c|c|l}
		Type & Abbreviation & Equations & Reference \\ \hline \hline
		Form of AV & vNRL & (\ref{eq:local_AV}) & \citet{VR50, L55} \\ \cline{2-4}
		& M97 & (\ref{eq:pairwise_AV}) & \citet{M97} \\ \hline
		& No & No & Do not use \\ \cline{2-4}
		Shock indicator & R+00 & (\ref{eq:MM97}) & \citet{R+00} \\ \cline{2-4}
		& CD10 & (\ref{eq:CD10b_st})-(\ref{eq:CD10b_to}) & \citet{CD10} \\ \hline
		& No & No & Do not use \\ \cline{2-4}
		Shear switch & B95 & (\ref{eq:B95}) & \citet{B95} \\ \cline{2-4}
		& CD10 & (\ref{eq:CD10s_st})-(\ref{eq:CD10s_to}) & \citet{CD10} \\ \hline
		$\vec{\nabla} \vec{\cdot} \vec{v}$ & SPH & (\ref{eq:SPH_div}) & \citet{L77} \\ \cline{2-4}
		& G+12 & (\ref{eq:G+12_div}) & \citet{G+12} \\ \hline
	\end{tabular}
	\label{tab:table1}
\end{table}

The Keplerian disk consists of cold gas orbiting around the central massive object.
The surface density and pressure of the disk is set to uniformly 1.0 and $10^{-6}$, which are the same as those used in \citet{H15}.
The centrifugal force $\vec{a}_\mathrm{ext}$ is added to each particle and given by the following equation:
\begin{eqnarray}
\vec{a}_\mathrm{ext} = - \frac{G M}{(|\vec{x}|^2 + \epsilon^2)^{3/2}} \vec{x}.
\end{eqnarray}
In this test we set $G = 1$ and $M = 1$ and $\epsilon$ is the softening length which prevents numerical overflow due to the particles which fall down close the center of the disk.
We set $\epsilon = 0.25$, if and only if $|\vec{x}| < 0.25$.

We constructed the initial particle distributions as concentric rings (see Appendix \ref{Sec:App2} for detail).
The inner and outer cutoff radii of the disk is $0.5$ and $2.0$.
We employ 46560 particles in total.
Note that we also tested the Cartesian grid initial distribution, similar to \citet{H15}.
The main mechanism for the decay of the disk is the same.
In this paper, therefore, we do not show the results for the Cartesian grid case.

In order to carry out the quantitive comparison of the results, we introduce rough estimate of the disk life time.
Let us consider the root mean square of the radius of the particles initially located at the inner edge of the disk.
When this value exceeds $10\%$ of the radius of the inner edge, we regard the disk as disrupted.

\begin{figure}[tb]
\plotone{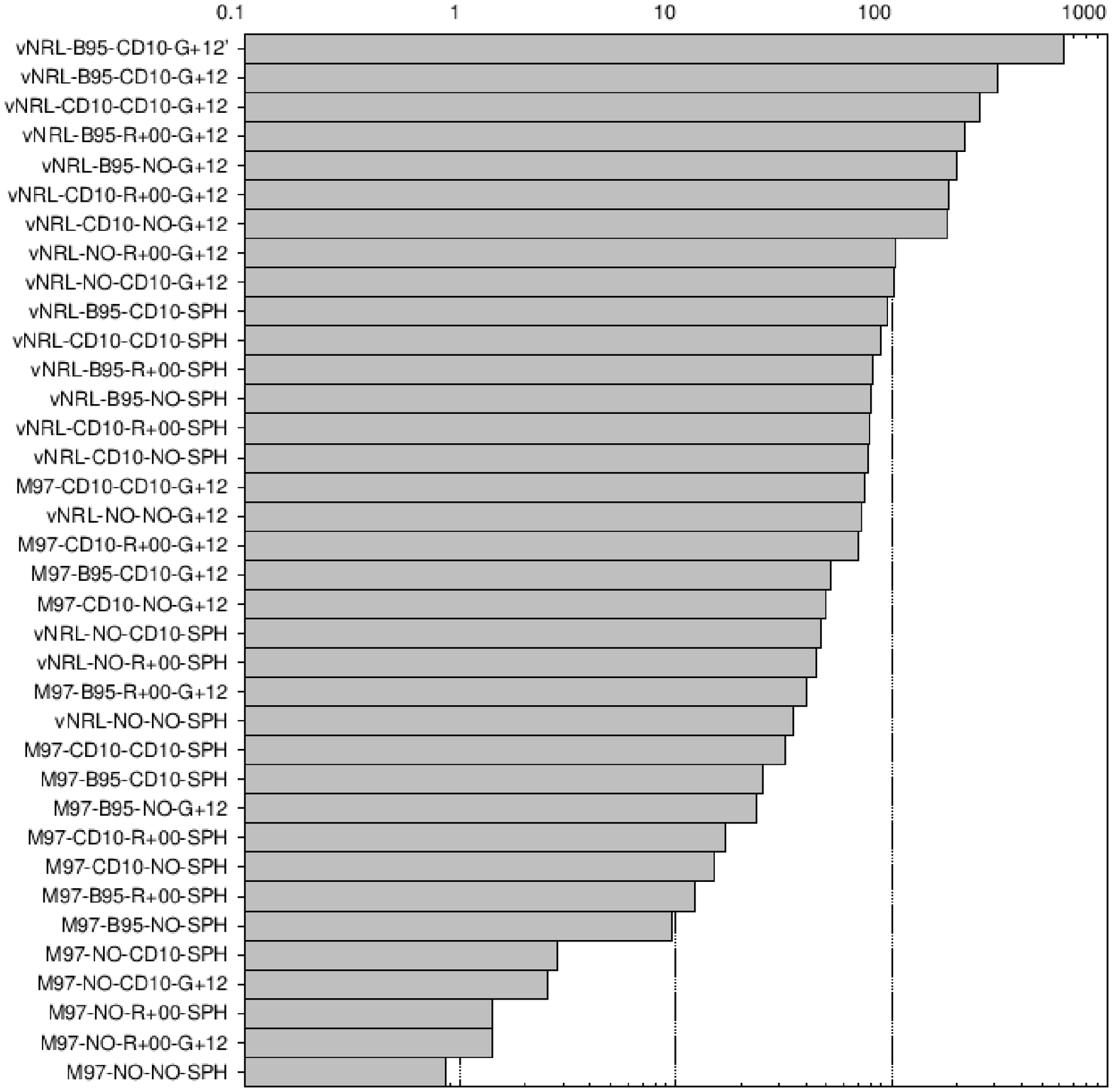}
\caption{
	Lifetime of the disk for each AV implementation in the unit of the Kepler time at $r = 0.5$ (inner edge of the disk).
	Note that the disk lifetime is shown in the log scale.
}
\label{fig:life_time}
\end{figure}

Figures \ref{fig:life_time} and \ref{fig:disk_ring1} show the lifetime and snapshots of the Keplerian disk for each AV implementation.
The differences among the lifetimes of the disks obtained with different implementations are very large.
In the case of vNRL-B95-CD10-G+12, the disk survived for the time more than 10 times longer than with the standard method used today (M97-B95-R+00-SPH).
Note that the most longest case, vNRL-B95-CD10-G+12$'$ indicates the results with vNRL-B95-CD10-G+12 scheme, but with $\alpha_\mathrm{min}^\mathrm{AV} = 0.025$ and $\alpha_\mathrm{max}^\mathrm{AV} = 0.5$.
These parameters in AV switch would also play important role for the long time evolution of the disk.

\begin{figure}
\plotone{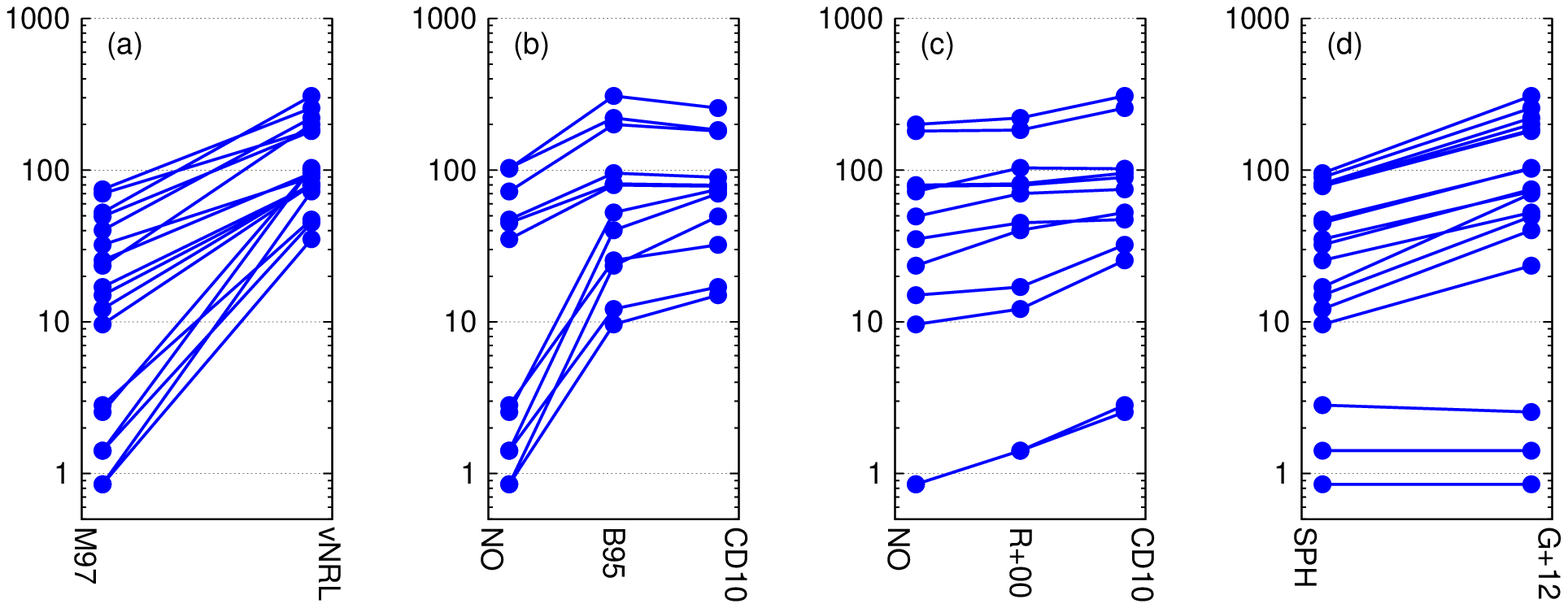}
\caption{
	The difference of the lifetime of the disk when (a)the AV scheme, (b)the shear switch, (c)the shock indicator and (d)derivative estimator are changed.
	Points connected by lines indicate the runs with the same method except for that shown in panel.
}
\label{fig:trend}
\end{figure}

To clarify the effects of each option to the disk lifetime, we show the dependences of the disk lifetime to each option in Fig. \ref{fig:trend}.
Here, we show how the lifetime of the disk changes when we change the method in one type, while kept the methods for other three types unchanged.
We can clearly see the following tendencies.
(i)The disk lifetime obtained with vNRL AV is much longer than that with the M97 AV.
(ii)Both of two shear switches extend the disk lifetime.
The CD10 switch works better than the B95 switch for M97 AV, while the B95 switch works better than the CD10 switch for the vNRL AV.
(iii)Both shock indicators extend the disk lifetime.
The CD10 shock indicator works better than R+00 does.
(iv)The G+12 derivative operator works better than the standard SPH derivative operator.
This might be due to the effect that the SPH derivative operator omits the surface term \citep[for detail, see Section 3.1 in][]{P08}.

\begin{figure}
\plotone{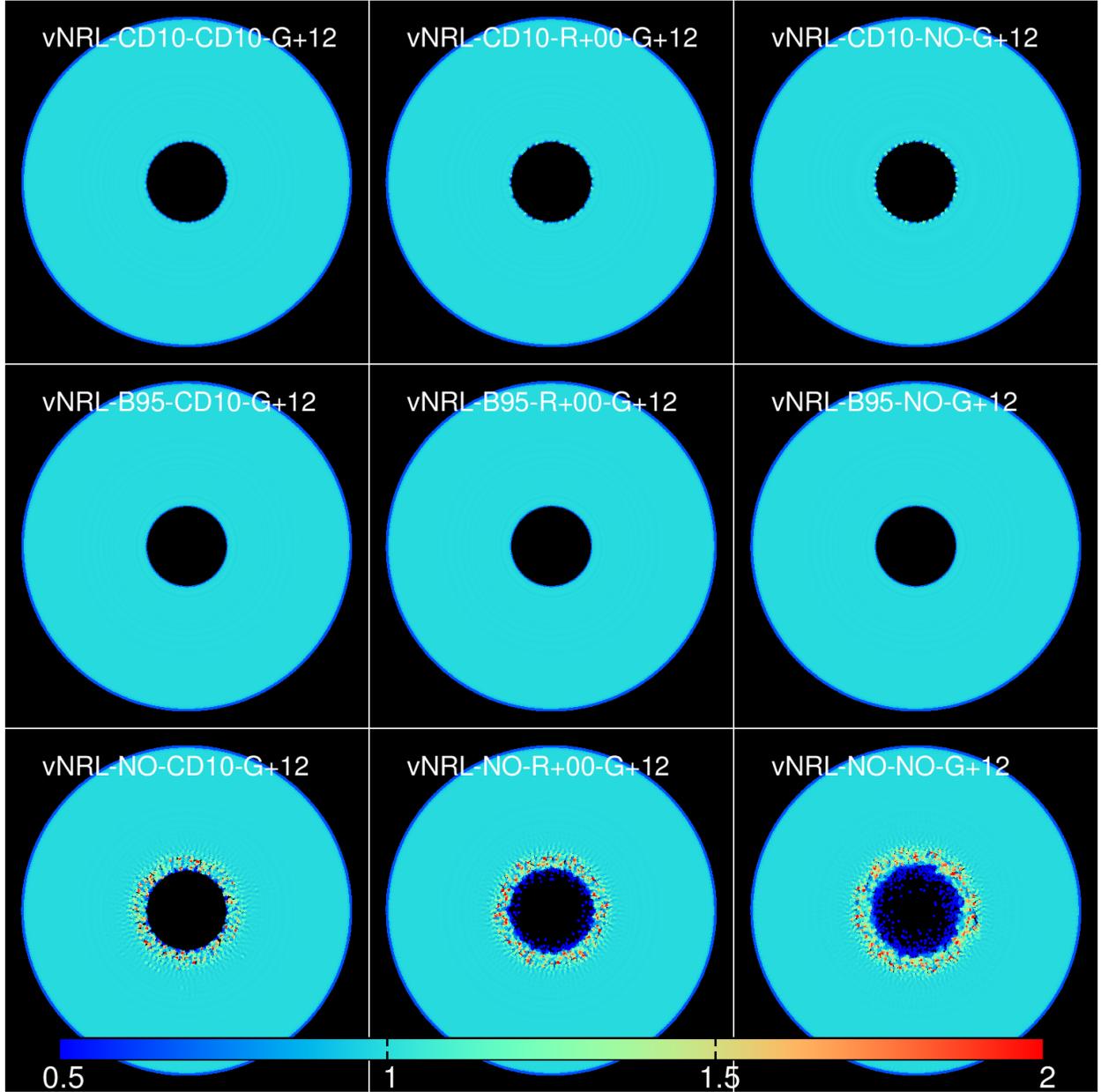}
\caption{
	Snapshots from the Keplerian disk test at $\sim 28$ orbits at $r = 0.5$ (inner edge of the disk) with each AV implementation.
	The color code of the density is given at the bottom.
	Since no disk life time with M97 AV exceeds $\sim 28$ orbits at $r = 0.5$, we only show the results of four cases.
}
\label{fig:disk_ring1}
\end{figure}

\begin{figure}
\figurenum{\ref{fig:disk_ring1}}
\plotone{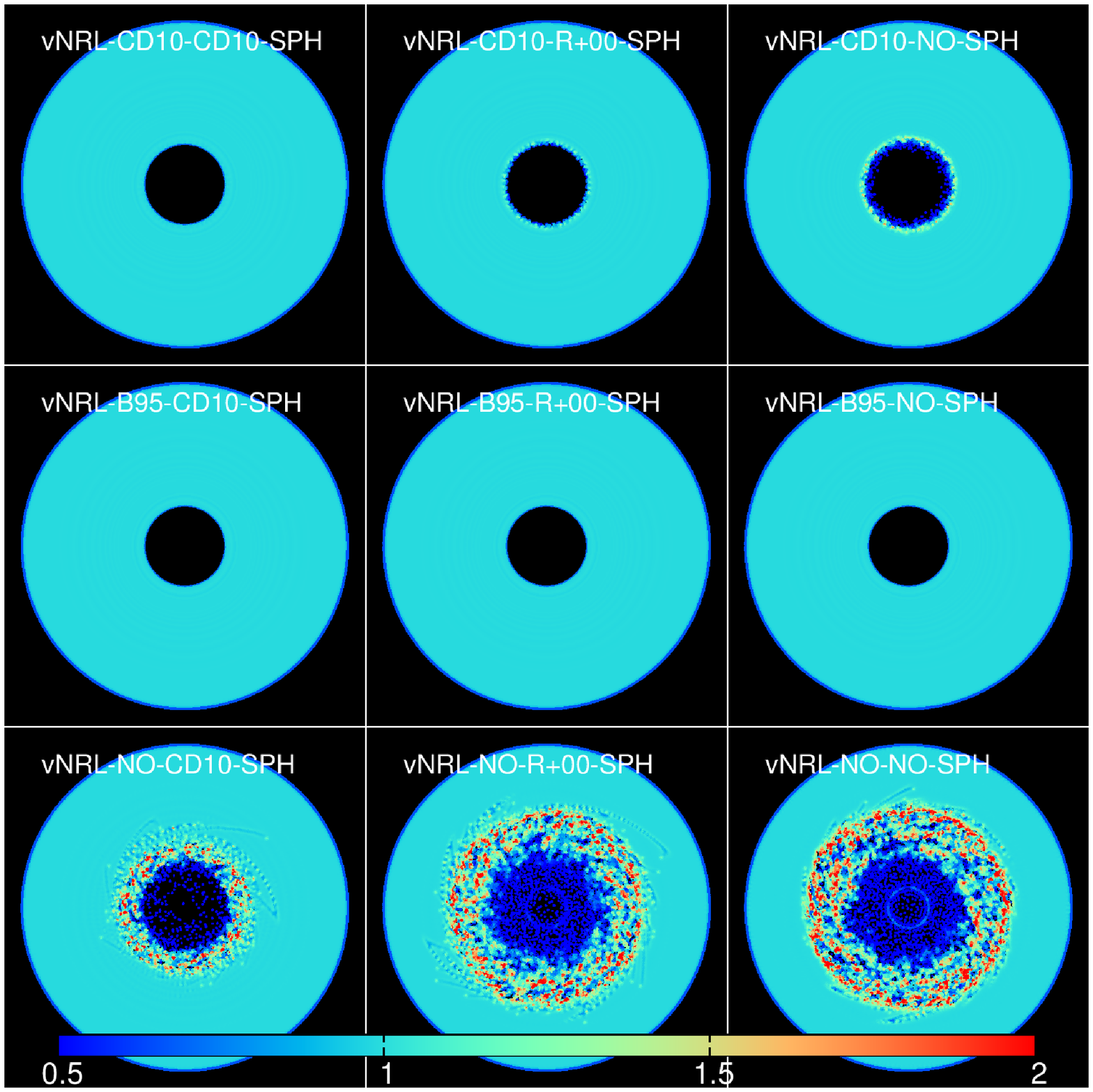}
\caption{
	Continued.
}
\label{fig:disk_ring2}
\end{figure}

\begin{figure}
\figurenum{\ref{fig:disk_ring1}}
\plotone{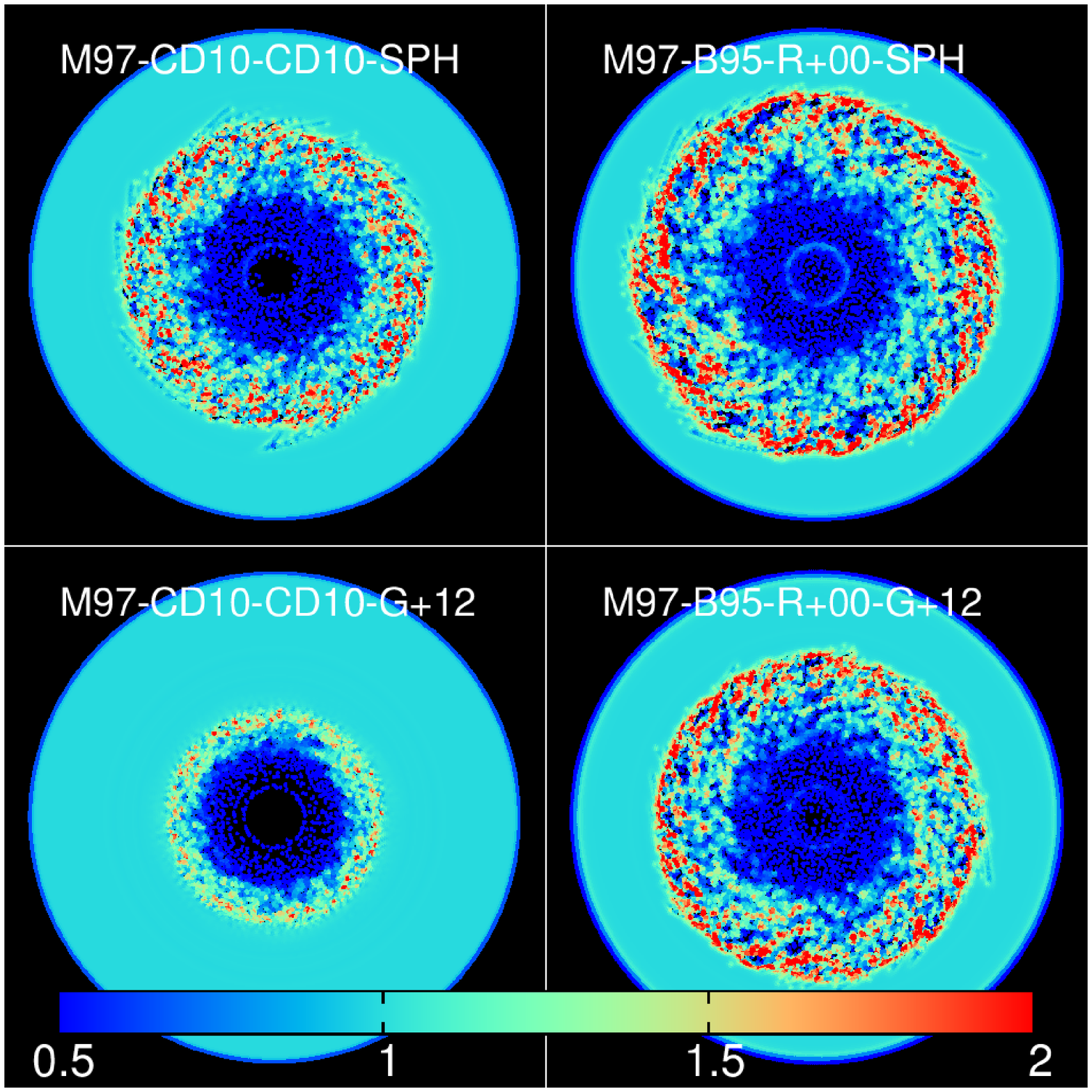}
\caption{
	Continued.
}
\label{fig:disk_ring3}
\end{figure}

\begin{figure}
\includegraphics[scale=0.9]{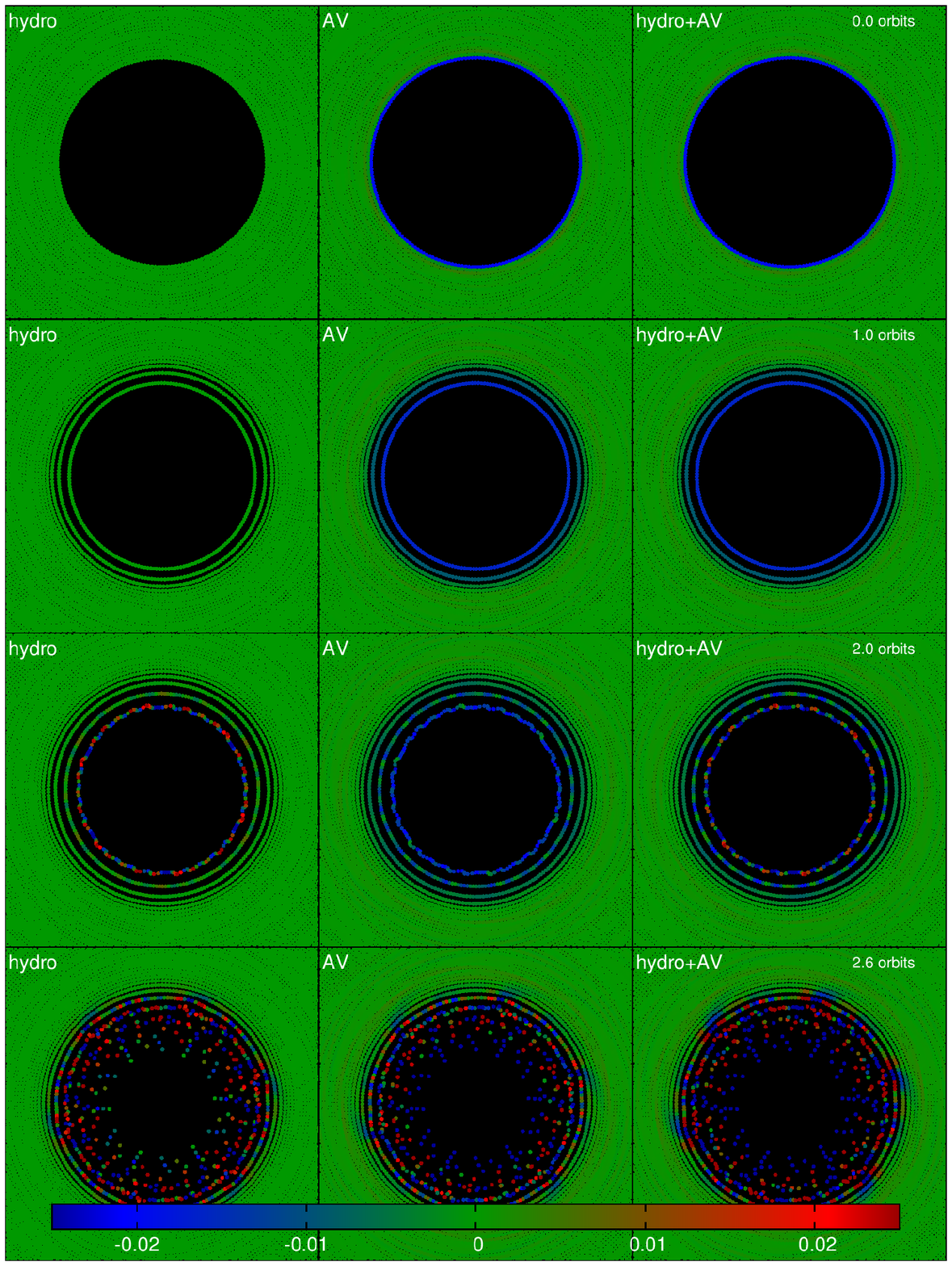}
\caption{
	The distribution of specific torque by hydrodynamical force, AV force and both of them from left to right for the run with M97-No-No-SPH.
	Only particles within $[0.75, 0.75]^2$ are shown.
	The color code indicates the torque.
	The snapshot times are normalized by the orbital time at the inner edge ($r = 0.5$) from top to bottom.
}
\label{fig:M97-NO-NO-SPH}
\end{figure}

\begin{figure}
\includegraphics[scale=0.9]{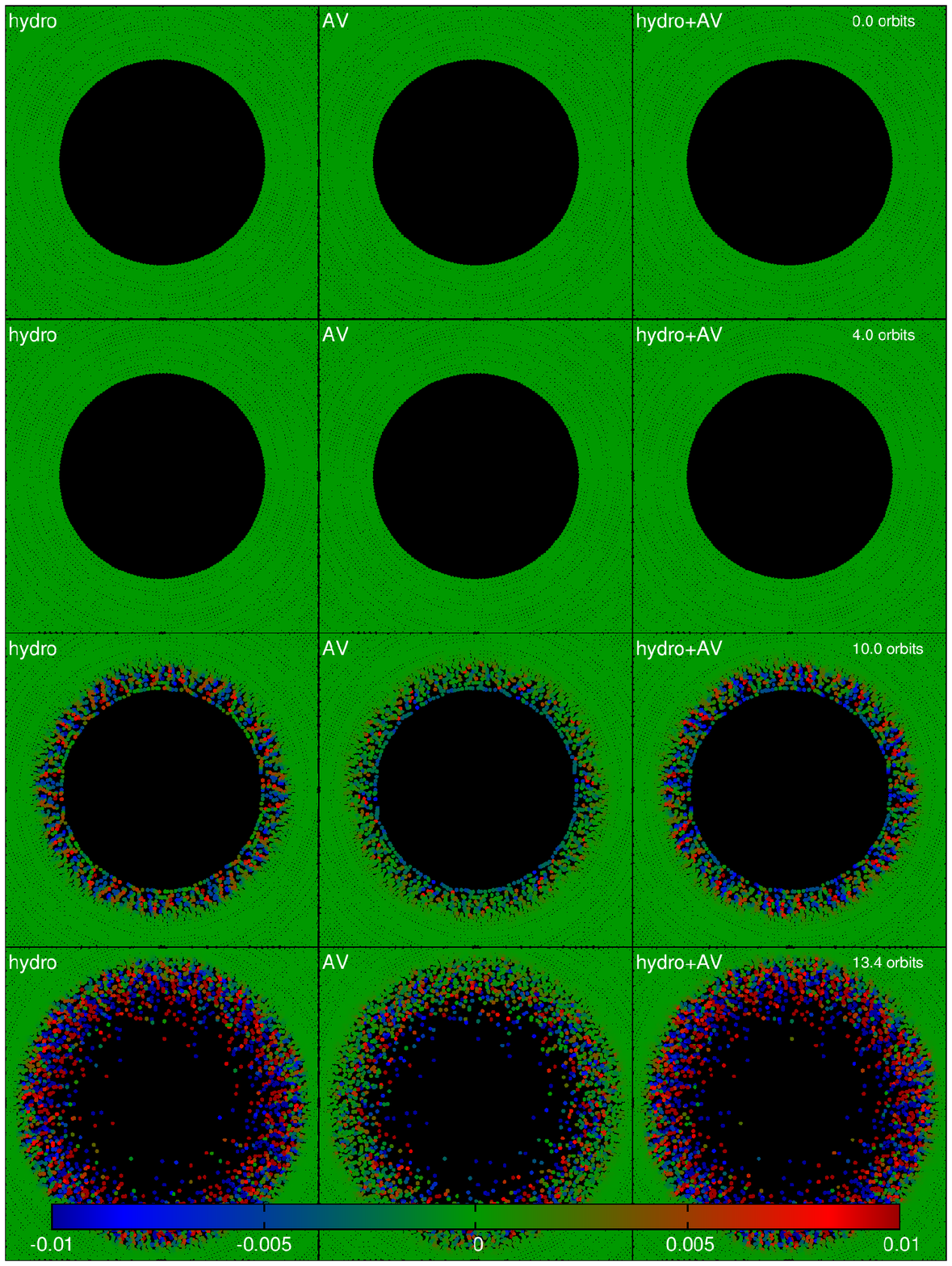}
\caption{
	The same as Fig. \ref{fig:M97-NO-NO-SPH}, but shows the results with M97-B95-R+00-SPH.
	Note that the time and the color bar are different.
}
\label{fig:M97-B95-R+00-SPH}
\end{figure}

\begin{figure}
\includegraphics[scale=0.9]{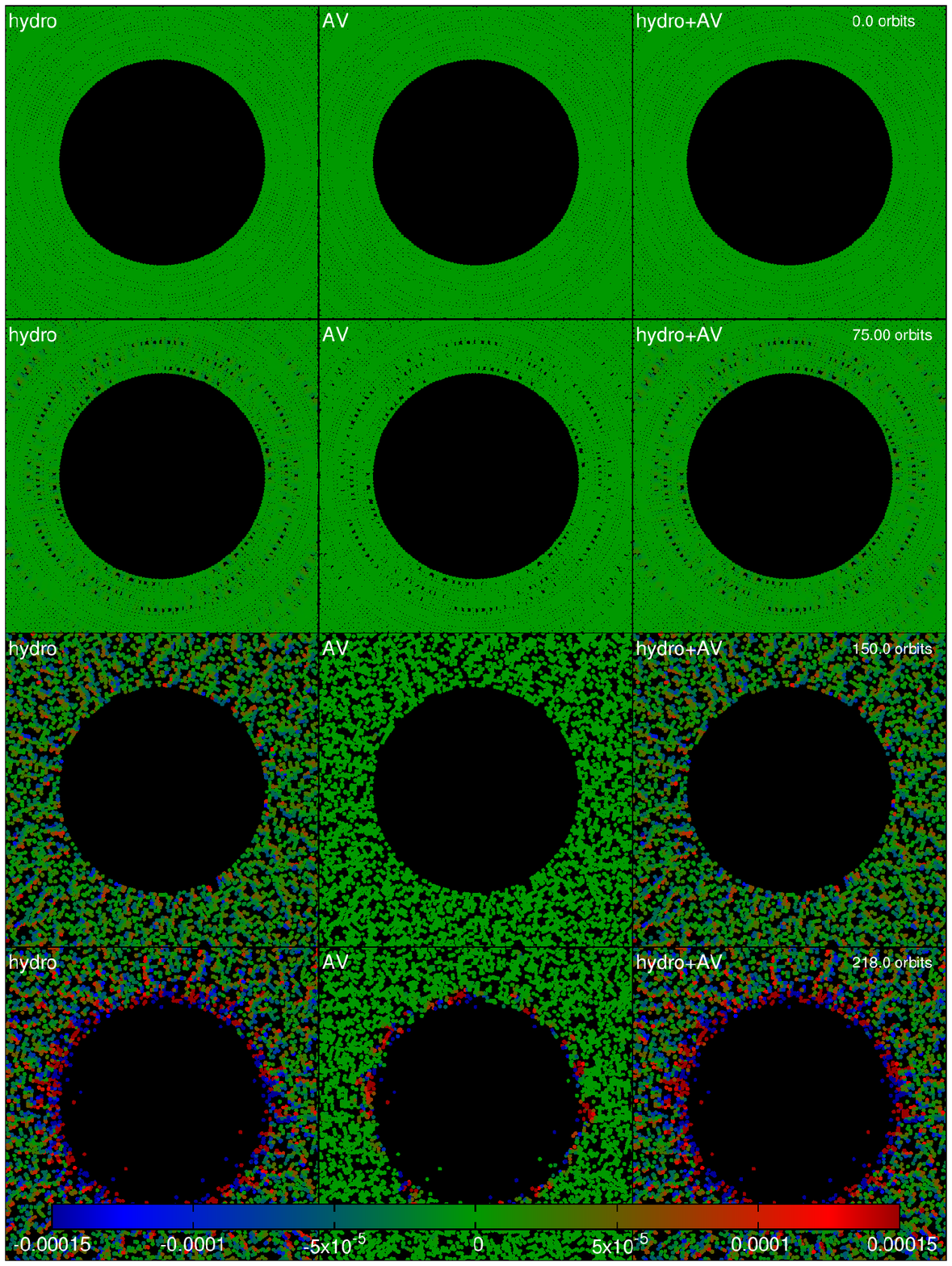}
\caption{
	The same as Fig. \ref{fig:M97-NO-NO-SPH}, but shows the results with vNRL-B95-R+00-G+12.
	Note that the time and the color bar are different.
}
\label{fig:vNRL-B95-R+00-G+12}
\end{figure}

\begin{figure}
\plotone{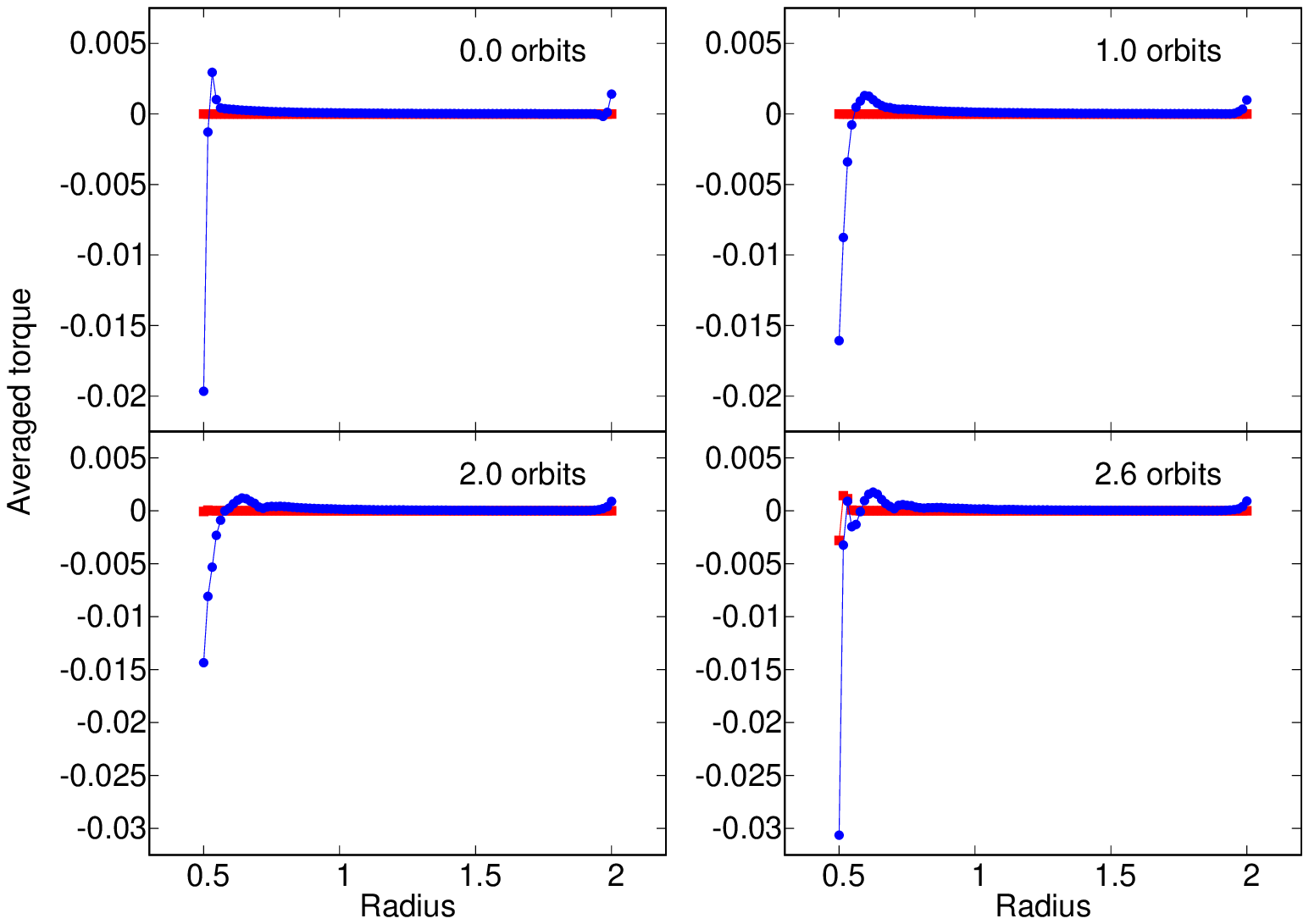}
\caption{
	The averaged torque for the run with M97-No-No-SPH plotted against the initial radius of particles.
	The blue points indicate the torque by AV and red points indicate those by hydrodynamical force.
	The times are normalized by the orbital time at the inner edge ($r = 0.5$).
}
\label{fig:M97-NO-NO-SPH_bin}
\end{figure}

\begin{figure}
\plotone{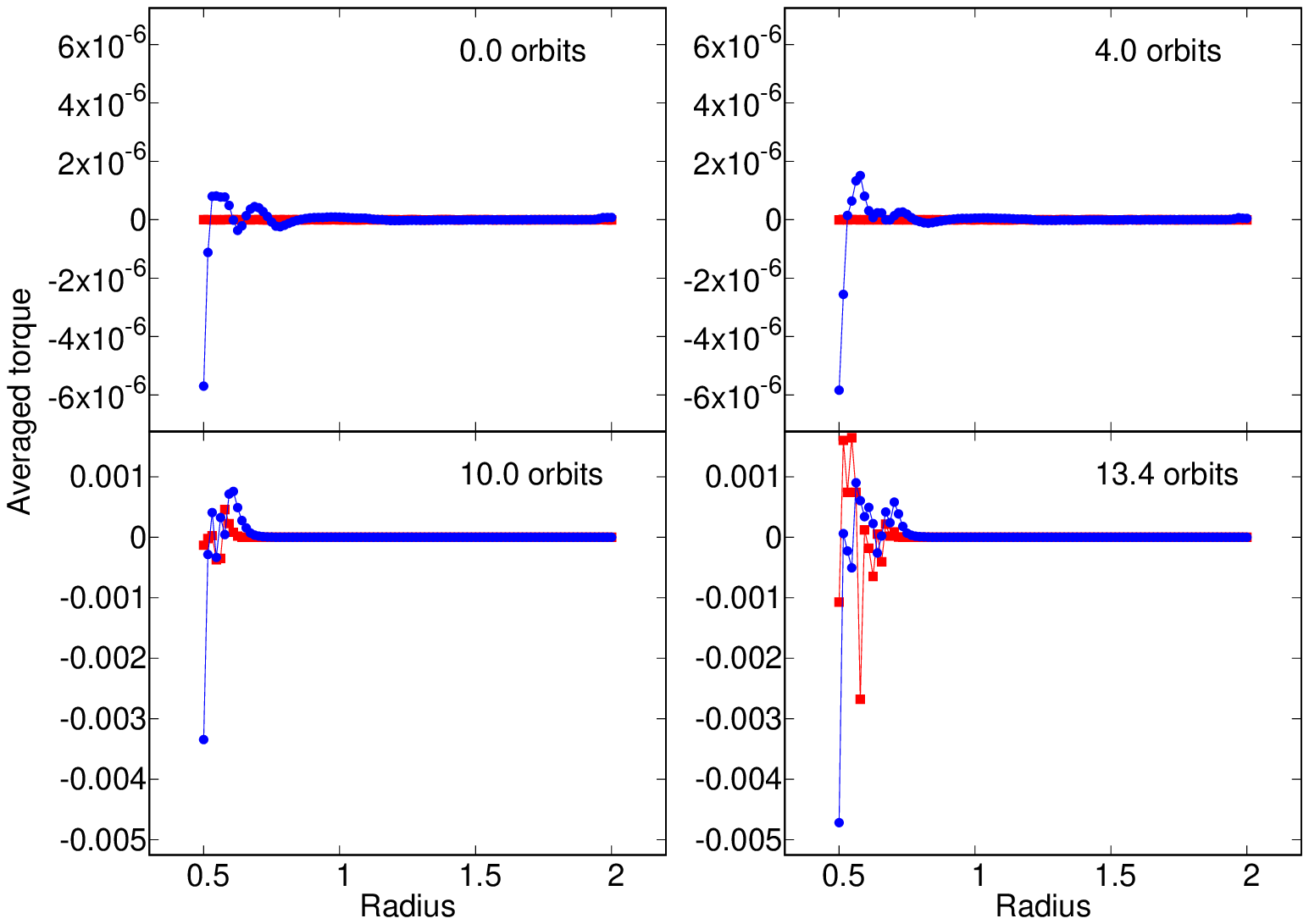}
\caption{
	Same as Fig. \ref{fig:M97-NO-NO-SPH_bin}, but for the run with M97-B95-R+00-SPH.
}
\label{fig:M97-B95-R+00-SPH_bin}
\end{figure}

\begin{figure}
\plotone{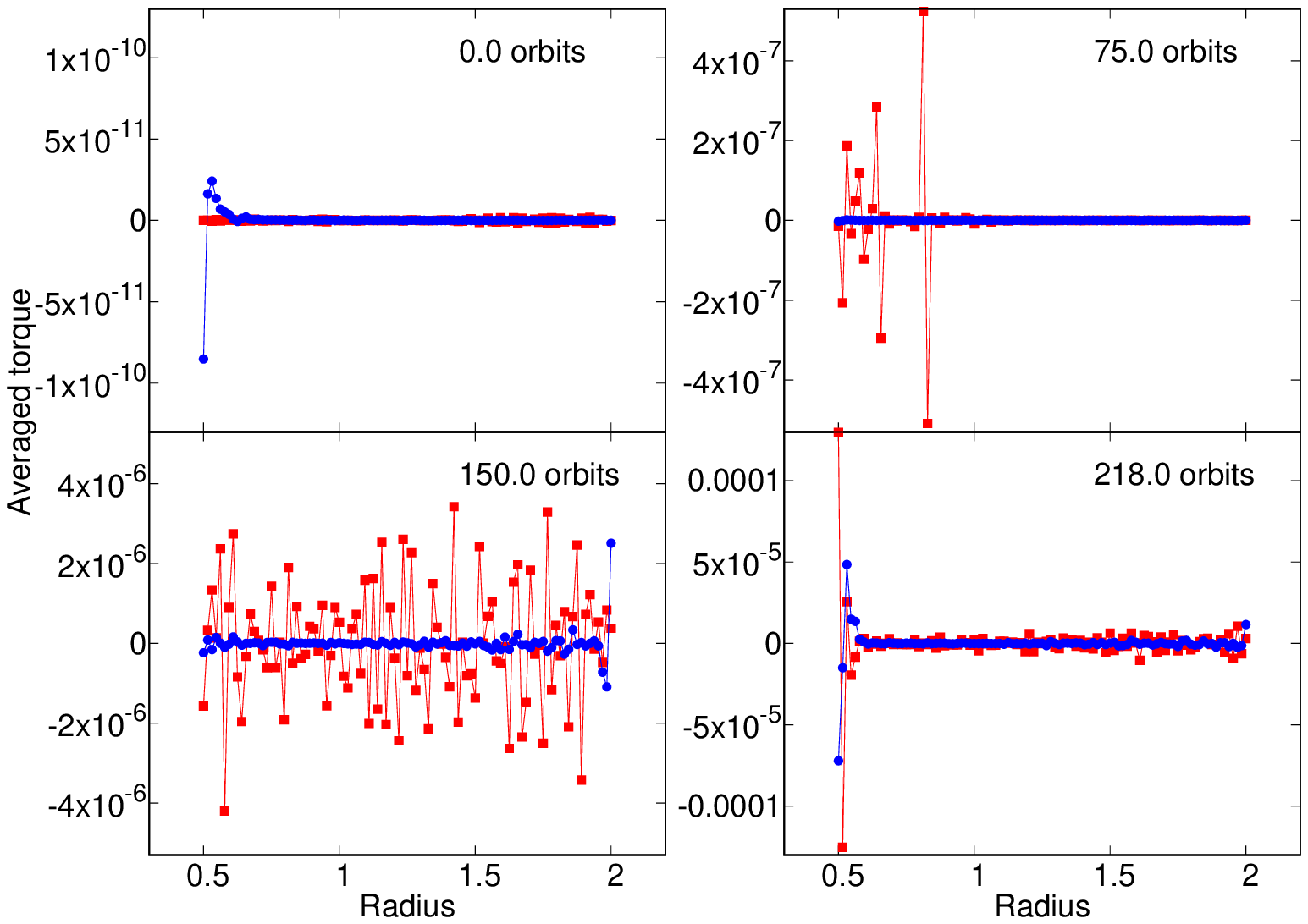}
\caption{
	Same as Fig. \ref{fig:M97-NO-NO-SPH_bin}, but for the run with vNRL-B95-R+00-G+12.
}
\label{fig:vNRL-B95-R+00-G+12_bin}
\end{figure}

Figures \ref{fig:M97-NO-NO-SPH}, \ref{fig:M97-B95-R+00-SPH} and \ref{fig:vNRL-B95-R+00-G+12} show the evolution of the inner part of the Keplerian disks for three representative runs (M97-No-No-SPH, M97-B95-R+00-SPH and vNRL-B95-R+00-G+12).
In Figs. \ref{fig:M97-NO-NO-SPH_bin}, \ref{fig:M97-B95-R+00-SPH_bin} and \ref{fig:vNRL-B95-R+00-G+12_bin} we show the hydrodynamical and AV torques as a function of the initial radius for these three runs.
There is theoretically no mechanism of the angular momentum transfer in the disks.
Hence the evolution of the disks is induced by the numerical effect.
These figures clearly describe how the disks break up.

With M97-No-No-SPH, at the very first step, a huge AV torque works at the inner edge (see Figs. \ref{fig:M97-NO-NO-SPH} and \ref{fig:M97-NO-NO-SPH_bin}).
This torque is negative and thus the inner edge of the disk is decelerated.
Consequently, the inner edge start to fall off from the rest of the disk.
After the innermost ring has become completely separated from the disk, the second innermost ring of particles starts to fall off and it is then followed by third ring, and then fourth, and so on.
Finally these isolated rings break up.
In Fig. \ref{fig:M97-NO-NO-SPH_bin}, we can clearly see that the AV torque to the innermost ring is initially very large and remains to be large.
Since we did not use any shear switch, M97 viscosity causes large drag to particles in the innermost ring, which is rotating faster than neighboring outer rings.

With M97-B95-R+00-SPH, the disk survived much longer (see Figs. \ref{fig:M97-B95-R+00-SPH} and \ref{fig:M97-B95-R+00-SPH_bin}).
We can see that the initial AV torque in Figs. \ref{fig:M97-B95-R+00-SPH} and \ref{fig:M97-B95-R+00-SPH_bin} is more than three orders of magnitude smaller than that in Figs. \ref{fig:M97-NO-NO-SPH} and \ref{fig:M97-NO-NO-SPH_bin}.
This much smaller AV torque resulted in the disk lifetime roughly ten times longer for M97-B95-R+00-SPH compared to that for M97-No-No-SPH.
Apparently, the way the disk is disrupted is quite different from that for the case of M97-No-No-SPH.
Inner region of the disk becomes disordered before the falling down of the innermost ring occurs, and from this disordered ring particles are eventually kicked out, and then complete break down of the inner region takes place.

With vNRL-B95-R+00-G+12, the disk survived for around $200$ orbital period at the inner edge (see Figs. \ref{fig:vNRL-B95-R+00-G+12} and \ref{fig:vNRL-B95-R+00-G+12_bin}).
At the beginning of the simulation, we can see from Fig. \ref{fig:vNRL-B95-R+00-G+12_bin} that the AV torque is of the orders of $10^{-11}$ which is five orders of magnitude smaller than that in the case of M97-B95-R+00-SPH.
The use of switches and the use of vNRL AV instead of the M97 AV reduce the AV torque drastically, resulting in the extension of the disk lifetime.

The disk finally breaks up after about $220$ orbits.
The mechanism of the breaking up of the disk seems to be similar to that in M97-B95-R+00-SPH.
The particle alignment becomes disordered first, and eventually particles are kicked out from the inner edge of the disk.
However, both the time to become disordered and the time to eject particles are much longer.
In figure \ref{fig:vNRL-B95-R+00-G+12} we can see that disorders of particles start to grow at $t = 75$ orbits.
Then, particles near the inner edge show perturbed motion which triggers the break up of the disk.
These disorders cause large unphysical hydrodynamical forces, as can be seen in Fig. \ref{fig:M97-B95-R+00-SPH_bin}.
As a result, the disorder propagates to the while disk.
Note that this azimuthal motion seems to have a mode with a wavenumber $6$.
This comes from the construction of the initial condition (see Appendix \ref{Sec:App2}).

Figure \ref{fig:inner} shows the time evolution of the AV torque per mass on the particles which are initially on the inner edge.
It is apparent that in the run with M97-No-No-SPH particles feel a large torque from the beginning to the end of the simulation.
The torque in the run with M97-B95-R+00-SPH is about three orders of magnitude smaller at the beginning, but it increases exponentially after $t \simeq 4$ until the break up of the disk.
The AV torque in the run with vNRL-B95-R+00-G+12 is much smaller than that with M97-B95-R+00-SPH.
The growth of the AV force is much slower for this case compared to the run with M97-B95-R+00-SPH.
It remains small until $t \sim 60$, and the exponential growth after $t \simeq 60$ is also slower.

\begin{figure}
\includegraphics[scale=1.0]{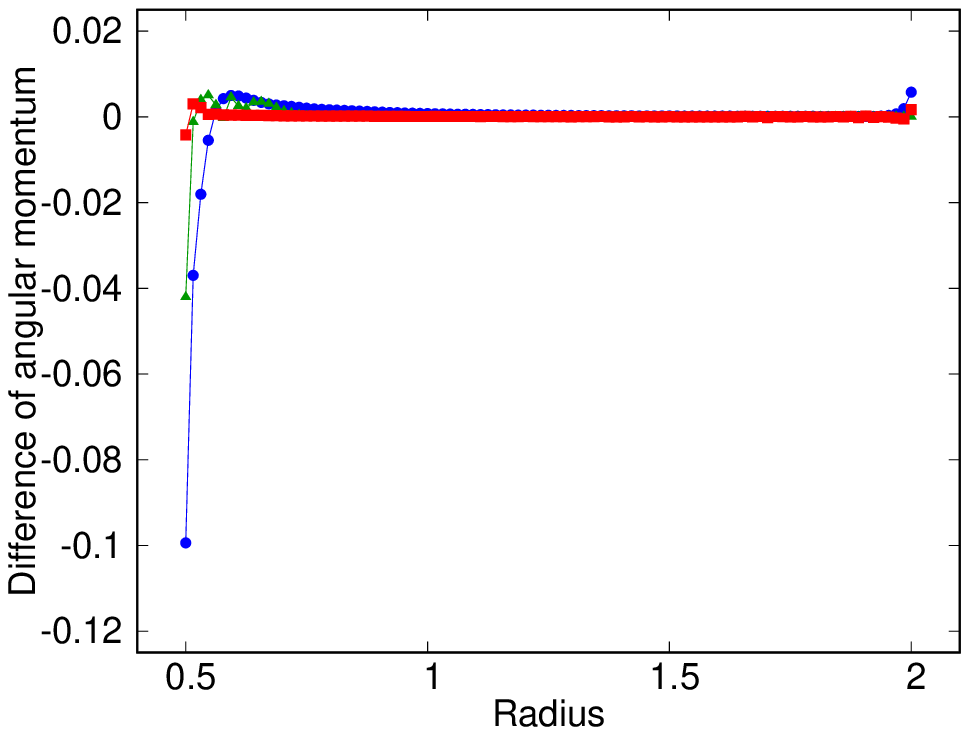}
\caption{
	Difference between the angular momentum at the initial step and those at the disk life time are plotted against the initial radius of particles.
	The red squares, green triangles and blue circles indicate the results of vNRL-B95-R+00-G+12 at $t = 300$ orbits, M97-B95-R+00-SPH at $t = 12$ orbits and M97-No-No-SPH at $t = 0.9$ orbits, respectively.
	The time is normalized in the orbital time at the inner edge.
}
\label{fig:AMtransport}
\end{figure}

Fig. \ref{fig:AMtransport} shows the difference between the radial distributions of angular momenta at the initial state at the end time.
We can see that unphysical transfer of angular momentum took place around inner and outer edges.
No such transfer can be seen in the bulk of the disk.
Near the inner edge, particles lose angular momenta while around the outer edge they gain angular momenta.
The magnitude of the angular momentum transfer at the inner edge is larger than those at the outer edge.
Thus, we conclude that the main reason for the break up of the disk is this angular momentum transfer at the inner edge.
Note that the transfer of angular momentum in the case of run with M97 AV is much faster than that in the case of vNRL AV.
For the M97 AV, the end of the run is at $t=0.9$, while for the vNRL AV, that is at $t=300$.

\begin{figure}
\plotone{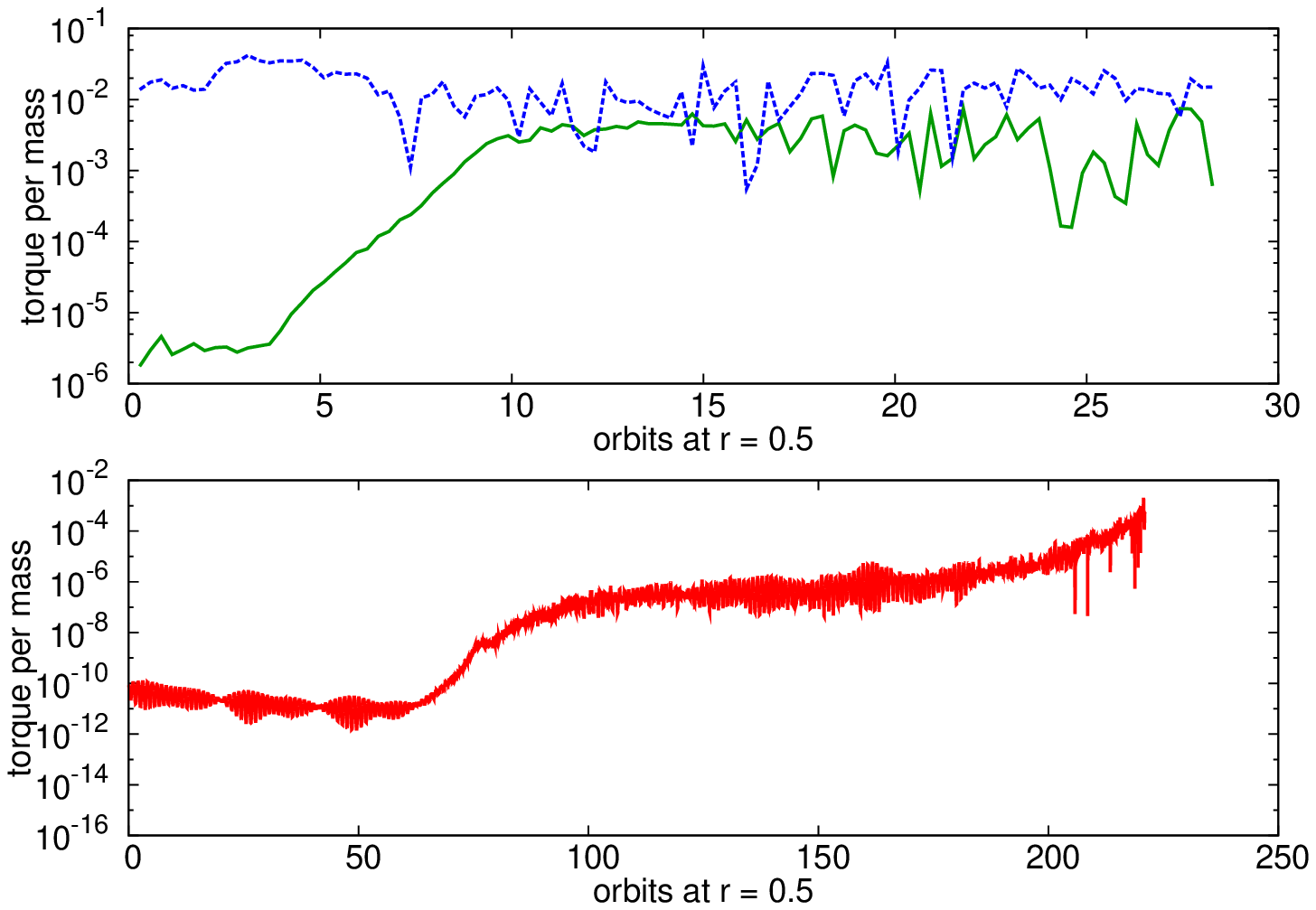}
\caption{
	Time vs. absolute value of the mean torque per mass due to AV exerted on the particles which are initially at the inner edge $r = 0.5$.
	The vertical axis is shown in the log scale.
	The dotted line and the solid line in the upper panel are the results for runs with M97-No-No-SPH and M97-B95-R+00-SPH, respectively.
	The lower panel shows shows the result for the run with vNRL-B95-R+00-G+12.
}
\label{fig:inner}
\end{figure}

\section{Summary}
It is well known that SPH has difficulties when used to simulate the evolution of cold and thin differentially rotating disks.
There are two possible reasons for this difficulty, namely, the error in the pressure gradient and spurious shear viscosity in AV.

In this paper, we present the result of a survey of the proposed implementations of AV, their switches and derivative operators.
For the simulation of cold, Keplerian disk, we found that vNRL AV gives much better result compared to widely used M97 AV.
Also, recently proposed G+12 operator for divergence calculation gives much better result compared to the traditional SPH divergence operator.
Note that a similar survey has been already done by \citet{H+14}.
However, they fixed the form of AV to the pairwise form.
Their results are consistent to our results with the pairwise AV.

Though the modern switches improve the behavior of the pair-wise AV significantly, we found the vNRL formulation behaves systematically better than the pair-wise AV.
The main reason for this difference is that the former responds to the compression while the latter responds to approaching pair of particles.
Thus, the pair-wise formulation cannot discriminate between the compression and shear.
This difference results in the difference between the results of Keplerian test.

One might think that vNRL AV cannot be used in the presence of the strong shock.
In appendix \ref{Sec:App3}, we present the results of the shock tube tests for vNRL and M97 AVs.
These tests show that vNRL AV does not lead to catastrophic results, although M97 AV shows better results.
Note that AV switches \citep[e.g.,][]{RH12, R15} can be easily combined with vNRL AV.
We conclude that vNRL AV can be possible alternative to the pairwise AV, especially if the system includes long time evolution and velocity shear.

\acknowledgments
\section*{acknowledgement}
We thank the anonymous referee for helpful comments.
Part of the research covered in this paper research was funded by MEXT program for the Development and Improvement for the Next Generation Ultra High-Speed Computer System, under its Subsidies for Operating the Specific Advanced Large Research Facilities.
This work was also supported in part by JSPS Grants-in-Aid for Scientific Research (Grant Number 26707007).

\appendix
\section{Equations of SPH}\label{Sec:App0}
In this section we briefly summarize the hydrodynamical part of SPH.
One of the most widely used formulation of SPH is the standard SPH (SSPH).
In this formulation, the equations of motion and energy are discretized as:
\begin{eqnarray}
\vec{a}_i & = & - \sum_j m_j \left[ \frac{p_i}{\Omega_i \rho_i^2} \vec{\nabla} W(\vec{x}_i - \vec{x}_j; h_i) + \frac{p_j}{\Omega_j \rho_j^2} \vec{\nabla} W(\vec{x}_i - \vec{x}_j; h_j) \right],\label{eq:motion}\\
\dot{u}_i & = & \frac{p_i}{\Omega_i \rho_i^2} \sum_j m_j (\vec{v}_{i} - \vec{v}_j) \vec{\cdot} \vec{\nabla} W(\vec{x}_i - \vec{x}_j; h_i).\label{eq:energy}
\end{eqnarray}
Note that since these equations are antisymmetric, the momentum and energy are conserved up to the machine epsilon.
The density and smoothing length are given by:
\begin{eqnarray}
\rho_i & = & \sum_j m_j W(\vec{x}_i - \vec{x}_j; h_i),\\
h_i & = & \eta \left( \frac{m_i}{\rho_i} \right)^{1/\nu},
\end{eqnarray}
where $\nu$ is the number of dimensions.
We set $\eta = 1.2$, unless otherwise specified.
Note that we need the equation of state to obtain the pressure from the density and specific internal energy.
In the following, we used the equation of state for ideal gas:
\begin{eqnarray}
p = (\gamma - 1) \rho u,
\end{eqnarray}
where $\gamma$ is the heat capacity ratio.
We set $\gamma$ to $1.4$, unless otherwise specified.
For the kernel function, we used the Wendland C${}_6$ kernel \citep{DA12}.

Recently, a novel formulation for the SPH, Density Independent SPH (DISPH) is proposed by \citet{SM13} to impose the description of the hydrodynamical instability of SSPH.
In this formulation, the equations of motion and energy are discretized as:
\begin{eqnarray}
\vec{a}_i & = & - (\gamma - 1) \sum_j m_j u_i u_j \left[ \frac{1}{\Omega_i q_i} \vec{\nabla} W(\vec{x}_i - \vec{x}_j; h_i) + \frac{1}{\Omega_j q_j} \vec{\nabla} W(\vec{x}_i - \vec{x}_j; h_j) \right], \label{eq:DImotion}\\
\dot{u}_i & = & (\gamma - 1) \frac{u_i}{\Omega_i q_i} \sum_j m_j u_j (\vec{v}_{i} - \vec{v}_j) \vec{\cdot} \vec{\nabla} W(\vec{x}_i - \vec{x}_j; h_i),\label{eq:DIenergy}
\end{eqnarray}
where
\begin{eqnarray}
q_i = \sum_j m_j u_j W(\vec{x}_i - \vec{x}_j; h_i).
\end{eqnarray}

In this paper, we employ the same time integrator to \citet{CD10}, which is the second order Runge-Kutta integrator scheme.
The procedure of this timestep integrator can be summarize as follows:\\
\step{1}\\
Drift $\vec{r}_i$:
\begin{eqnarray}
\vec{r}_i^{(n+1)} & = & \vec{r}_i^{(n)} + \vec{v}_i^{(n)} \Delta t + \vec{a}_i^{(n)} \frac{\Delta t^2}{2},
\end{eqnarray}
where the superscript $(n)$ indicates the value of $n$-th step.\\
\step{2}\\
Predict the velocity and energy at next step:
\begin{eqnarray}
\vec{v}_{i; \mathrm{pred}}^{(n+1)} & = & \vec{v}_i^{(n)} + \vec{a}_i^{(n)} \Delta t,\\
u_{i; \mathrm{pred}}^{(n+1)} & = & u_i^{(n)} + \dot{u}_i^{(n)} \Delta t.
\end{eqnarray}
At this step, we calculate the $\vec{\nabla} \vec{\cdot} \vec{v}_i^{(n)}$, $\vec{\nabla} \vec{\times} \vec{v}_i^{(n)}$ and $\vec{\nabla} \vec{\otimes} \vec{v}_i^{(n)}$ to obtain $\alpha_i^{\mathrm{AV}(n+1)}$ and $f_i$.
We then compute $\vec{a}_i^{(n+1)}$ and $u_i^{(n+1)}$ by using $\vec{v}_{i; \mathrm{pred}}^{(n+1)}$ and $u_{i; \mathrm{pred}}^{(n+1)}$.\\
\step{3}\\
Correct $\vec{v}_i$ and $u_i$:
\begin{eqnarray}
\vec{v}_i^{(n+1)} & = & \vec{v}_i^{(n)} +  \left( \vec{a}_i^{(n)} + \vec{a}_i^{(n+1)} \right) \frac{\Delta t}{2},\\
u_i^{(n+1)} & = & u_i^{(n)} + \left( \dot{u}_i^{(n)} + \dot{u}_i^{(n+1)} \right) \frac{\Delta t}{2}.
\end{eqnarray}
Note that in the case the acceleration depends on only the position and not on the velocity (e.g., gravity), Step2 can be omitted.
In this case, this integrator results in the second order leapfrog integrator.

\section{The derivation of the derivative operator of G+12}\label{Sec:App1}
Let us consider the expansion of an arbitrary physical quantity $A(\vec{x}')$ around $\vec{x}$:
\begin{eqnarray}
A(\vec{x}') = A(\vec{x}) + (\vec{x}' - \vec{x}) \vec{\cdot} \vec{\nabla} A(\vec{x}) + \mathcal{O}\left( (\vec{x}' - \vec{x})^2 \right) \label{eq:Taylor}.
\end{eqnarray}
Multiplying $(\vec{x}' - \vec{x})$ to Eq. (\ref{eq:Taylor}) yields
\begin{eqnarray}
\left[ A(\vec{x}') - A(\vec{x}) \right] (\vec{x}' - \vec{x}) & = & (\vec{x}' - \vec{x}) \left[ (\vec{x}' - \vec{x}) \vec{\cdot} \vec{\nabla} A(\vec{x}) \right],\\
& = & \left[ (\vec{x}' - \vec{x}) \vec{\otimes} (\vec{x}' - \vec{x}) \right] \vec{\nabla} A(\vec{x}).
\end{eqnarray}
Taking the convolution of both sides and applying SPH discretization, we obtain
\begin{eqnarray}
\sum_j \left( A_j - A_i \right) (\vec{x}_j - \vec{x}_i) \frac{m_j}{\rho_j} W(\vec{x}_i - \vec{x}_j; h_i) = \mat{M}_i \vec{\nabla} A_i, \\
\mat{M}_i = \sum_j (\vec{x}_j - \vec{x}_i) \vec{\otimes} (\vec{x}_j - \vec{x}_i) \frac{m_j}{\rho_j} W(\vec{x}_i - \vec{x}_j; h_i),
\end{eqnarray}
and
\begin{eqnarray}
\vec{\nabla} A_i = \sum_j \left( A_j - A_i \right) \mat{M}^{-1} (\vec{x}_j - \vec{x}_i) \frac{m_j}{\rho_j} W(\vec{x}_i - \vec{x}_j; h_i).
\end{eqnarray}

It is easy and straightforward to extend this equation to a vector field.
Let us consider the derivative of $\alpha$-component of vector $\vec{v}$ by $\beta$-component:
\begin{eqnarray}
\nabla^\beta v_i^\alpha & = & \sum_j \left( v_j^\alpha - v_i^\alpha \right) G_{ji}^\beta, \\
\vec{G}_{ji} & = & \frac{m_j}{\rho_j} \mat{M}_i^{-1} (\vec{x}_j - \vec{x}_i) W(\vec{x}_i - \vec{x}_j; h_i).
\end{eqnarray}
The divergence of \vec{v} is then given by
\begin{eqnarray}
\vec{\nabla} \vec{\cdot} \vec{v}_i & = & \sum_\delta \nabla^\delta v_i^\delta\\
& = & \sum_\delta \sum_j \left( v_j^\delta - v_i^\delta \right) G_{ji}^\delta, \\
& = & \sum_j \left( \vec{v}_j - \vec{v}_i \right) \vec{\cdot} \vec{G}_{ji}.
\end{eqnarray}
Similarly, we can easily derive the expression for the rotation and dyadic;
\begin{eqnarray}
\vec{\nabla} \vec{*} \vec{v}_i = \sum_j \left( \vec{v}_j - \vec{v}_i \right) \vec{*} \left[ \mat{M}^{-1} (\vec{x}_j - \vec{x}_i) \right] \frac{m_j}{\rho_j} W(\vec{x}_i - \vec{x}_j; h_i),
\end{eqnarray}
where $\vec{*}$ is a placeholder operator for $\vec{*} \in \{\vec{\cdot}, \vec{\times}, \vec{\otimes}\}$.

\section{Initial particle placement}\label{Sec:App2}
For the initial distribution of particles, we used concentric rings.
\citet{SC97} made concentric rings by converting a unit square into a circle.
In this section, following \citet{SC97} to realize a uniform density disk, place particles on concentric rings.
These rings are obtained by simple coordinate transformation from regular particle placement in regular $n$-gon.
The procedure of this algorithm is follows, for the case of a circle with unit radius expressed by $N$ rings.
Repeat the following for $1 \leq i \leq N$.
Place $n \times i$ the particles in the circle of radius $r_i = i / N$ with equal spacing in azimuthal direction.
In this paper, we use $n = 6$.
Note that this procedure can be extended to disks with non-uniform surface density profile.
We will report it in forthcoming paper.

We can start the calculation from this initial placement.
However, we found that particles on the inner edge show radial oscillation which shortens the disk lifetime significantly.
Thus, before performing the numerical simulations, we stabilize the initial conditions by adding the damping term to the radial component of acceleration;
\begin{eqnarray}
\vec{a}^\mathrm{damp}_i = - \frac{C^\mathrm{damp}}{\Delta t} \frac{\vec{x}_i \vec{\cdot} \vec{v}_i}{|\vec{x}_i|^2} \vec{x}_i,
\end{eqnarray}
where $C^\mathrm{damp}$ is set to $0.1$.
Note that in this process we ignore the AV acceleration.
Time of process is set to $10 \times 2 \pi$.
We used the final snapshot of this process as the initial condition of the calculations done in Sec. \ref{Sec:test}.

\section{1D and 2D shock wave tests}\label{Sec:App3}
The shock tube problem is one of the most commonly used test problems to check the capability of numerical methods to handle the shocks.
In this section we present the results of the shock tube test for both SSPH and DISPH.
We used a 1D computational domain $[-0.5:0.5)$.
Here we performed two shock tube tests; one is the standard Sod shock tube test \citep{S79} and the other includes strong shock.
The initial condition of former test is given by:
\begin{eqnarray}
(\rho, p, v) = \left\{ \begin{array}{ll}
	(1, 1, 0) & (x < 0), \\
	(0.5, 0.2, 0) & (\mathrm{otherwise}).
\end{array} \right.
\end{eqnarray}
Tnitial condition for the strong shock test is given by:
\begin{eqnarray}
(\rho, p, v) = \left\{ \begin{array}{ll}
	(1, 1000, 0) & (x < 0), \\
	(1, 0.001, 0) & (\mathrm{otherwise}).
\end{array} \right.
\end{eqnarray}
We place 768 equal mass particles for the standard shock tube test and 1024 equal mass particles for the strong shock test.
For the strong shock test, we set $\eta = 1.6$.

\begin{figure}
\plotone{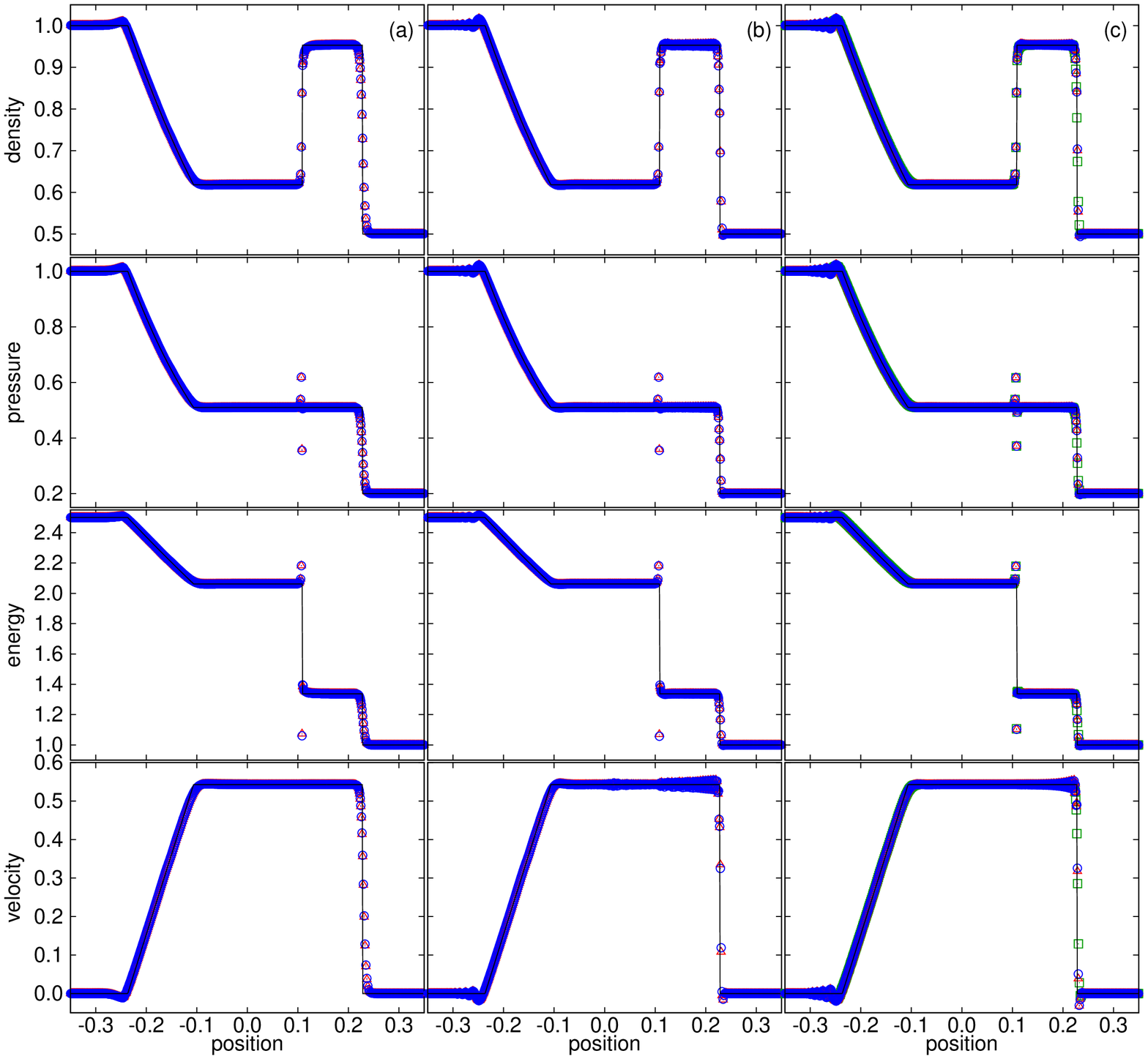}
\caption{
	Density, pressure, specific internal energy and velocity for the shock tube test at $t = 0.2$ with standard SPH.
	In the left panel, the red triangles indicate the results with vNRL-No-No-G+12, while blue circles indicate those with vNRL-No-No-SPH.
	In the central panel, the red triangles indicate the results with vNRL-No-R+00-G+12, while blue circles indicate those with vNRL-No-R+00-SPH.
	In the right panel, the red triangles indicate the results with M97-No-R+00-G+12, blue circles indicate those with M97-No-R+00-SPH and the green squares indicate those with M97-No-No-SPH.
	The solid curve indicates the analytic solution.
}
\label{fig:shocktube_S}
\end{figure}

Figure \ref{fig:shocktube_S} shows the results of the standard shock tube test.
All runs show good agreement with the analytic solution.
However, vNRL AV produces somewhat broader shock than M97-AV does (Figs. \ref{fig:shocktube_S}a and \ref{fig:shocktube_S}c).
The R+00 switch narrows the shock width for vNRL AV (Fig. \ref{fig:shocktube_S}b).
However, in the post shock region, wiggles can be seen, especially in the velocity.
The difference between SPH derivative operator and G+12 derivative operator are small (see blue circles and red triangles in Fig. \ref{fig:shocktube_S}).

\begin{figure}
\plotone{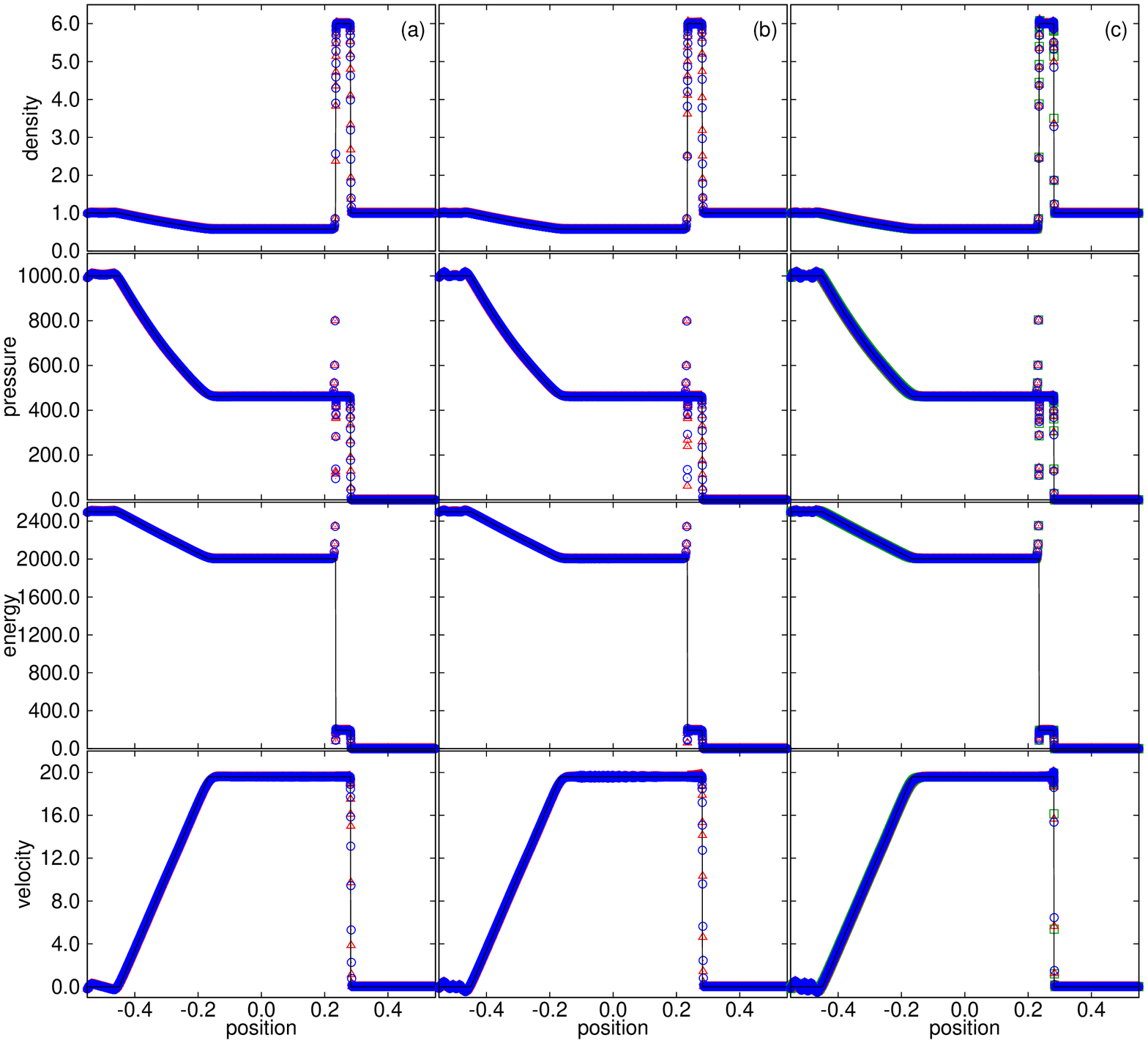}
\caption{
	The same as Fig. \ref{fig:shocktube_S}, but for the strong shock test.
}
\label{fig:strongshock_S}
\end{figure}

Figure \ref{fig:strongshock_S} shows the results of the strong shock test.
Similar trend to the standard shock tube test can be seen.
No significant difference can be seen between the results with vNRL AV and M97 AV.
However, unlike the standard shock case, vNRL AV produces broader shock, even when the R+00 switch is used.

\begin{figure}
\plotone{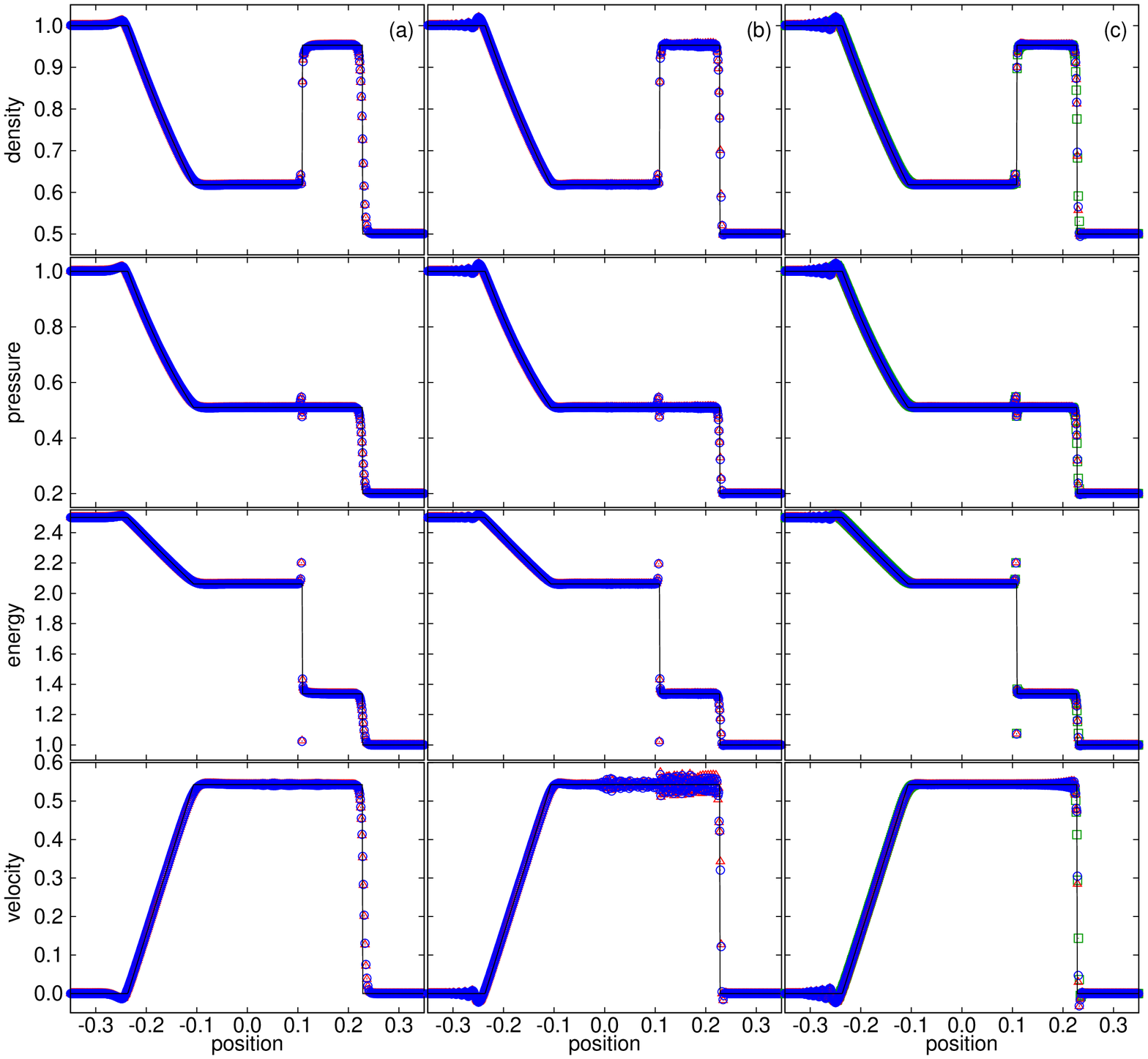}
\caption{
	The same as Fig. \ref{fig:shocktube_S}, but with DISPH.
}
\label{fig:shocktube_DI}
\end{figure}

\begin{figure}
\plotone{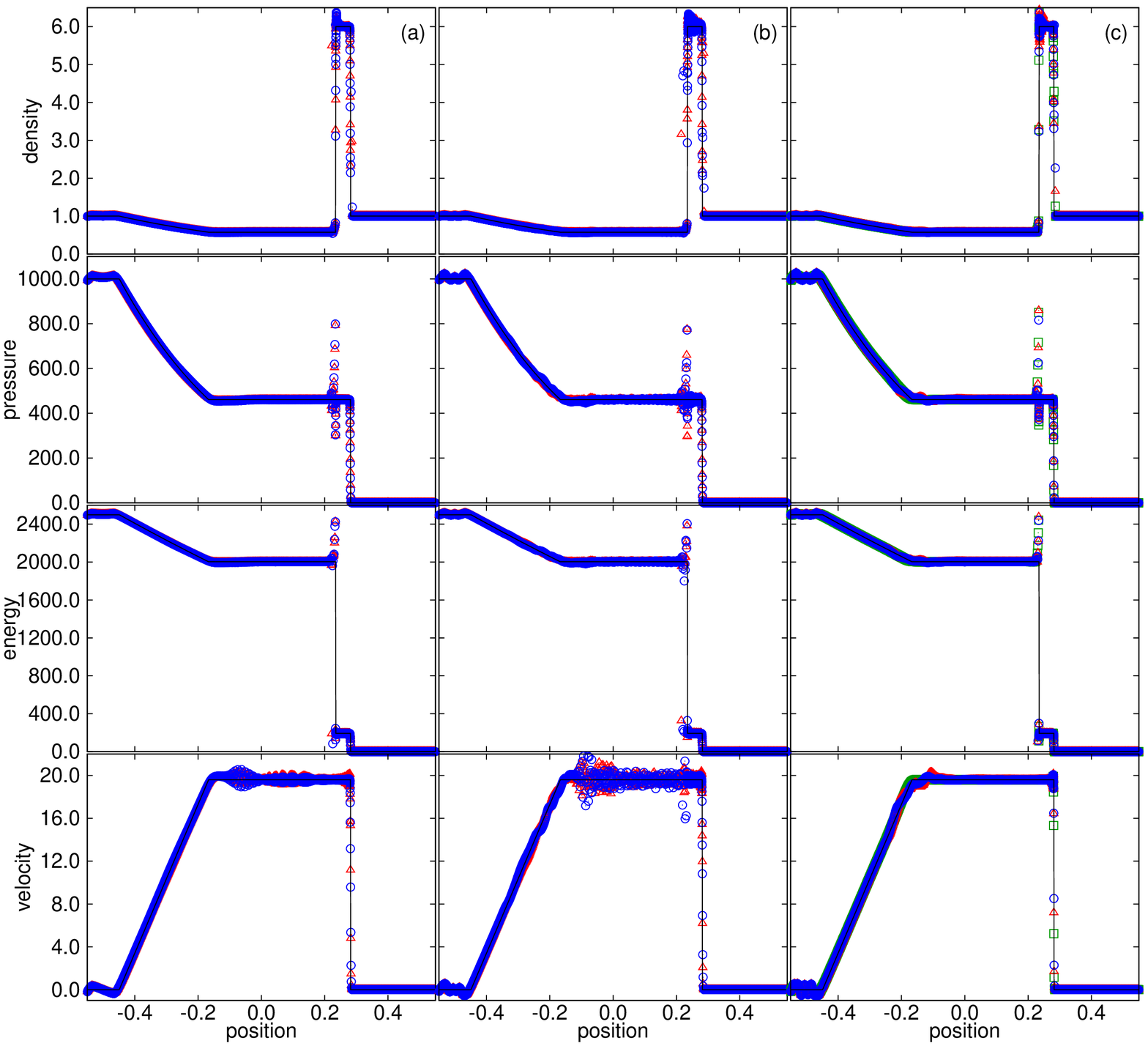}
\caption{
	The same as Fig. \ref{fig:strongshock_S}, but with DISPH.
}
\label{fig:strongshock_DI}
\end{figure}

Figures \ref{fig:shocktube_DI} and \ref{fig:strongshock_DI} show the same as Figs. \ref{fig:shocktube_S} and \ref{fig:strongshock_S} but with DISPH.
With standard shock tube test, DISPH shows similar trend to SSPH, while with strong shock test, DISPH tends to produce somewhat noisy post shock velocity.
This is because DISPH assumes that the pressure is smooth, although in strong shock test the initial pressure difference is very large.
We, however, note that there is a prescription for the strong shock with DISPH.
In \cite{SM13}, slightly modified DISPH, which does not smooth pressure but smooth the power of pressure, is proposed.
This can improve the accuracy of strong shock with DISPH.

We also performed the 2D Noh cylindrical implosion test, which involves strong shock \citep{N87}.
In this test, we consider the two dimensional computational domain ($-0.5 < x < 0.5$ and $-0.5 < y < 0.5$) filled with a fluid.
Initially, the density and pressure of the fluid in the domain have uniform values and they are set to unity and $10^{-6}$.
The initial velocity of the fluid is set to the magnitude $1$ with the direction toward the origin of coordinates.
The heat capacity ratio is set to $5/3$.
We construct the initial particle distribution from the same way to the Keplerian ring test (Appendix \ref{Sec:App2}).
The radial separation between each ring is set to $1/128$.
The analytic density solutions is
\begin{eqnarray}
\rho(t) = \left\{ \begin{array}{ll}
	16 & \left(r < \displaystyle\frac{t}{3} \right), \\
	1 + \displaystyle\frac{t}{r} & (\mathrm{otherwise}).
\end{array} \right.
\end{eqnarray}

\begin{figure}
\plotone{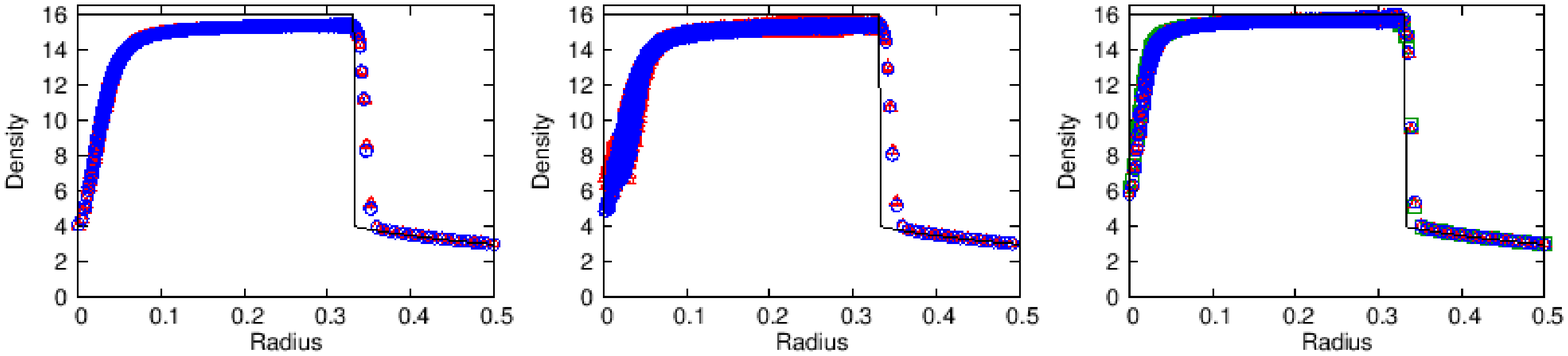}
\caption{
	Radius vs. density profiles of Noh implosion test are shown.
	In the left panel, the red triangles indicate the results with vNRL-No-No-G+12, while blue circles indicate those with vNRL-No-No-SPH.
	In the central panel, the red triangles indicate the results with vNRL-No-R+00-G+12, while blue circles indicate those with vNRL-No-R+00-SPH.
	In the right panel, the red triangles indicate the results with M97-No-R+00-G+12, blue circles indicate those with M97-No-R+00-SPH and the green squares indicate those with M97-No-No-SPH.
	The solid curve indicates the analytic solution.
}
\label{fig:Noh_2D}
\end{figure}

\begin{figure}
\plotone{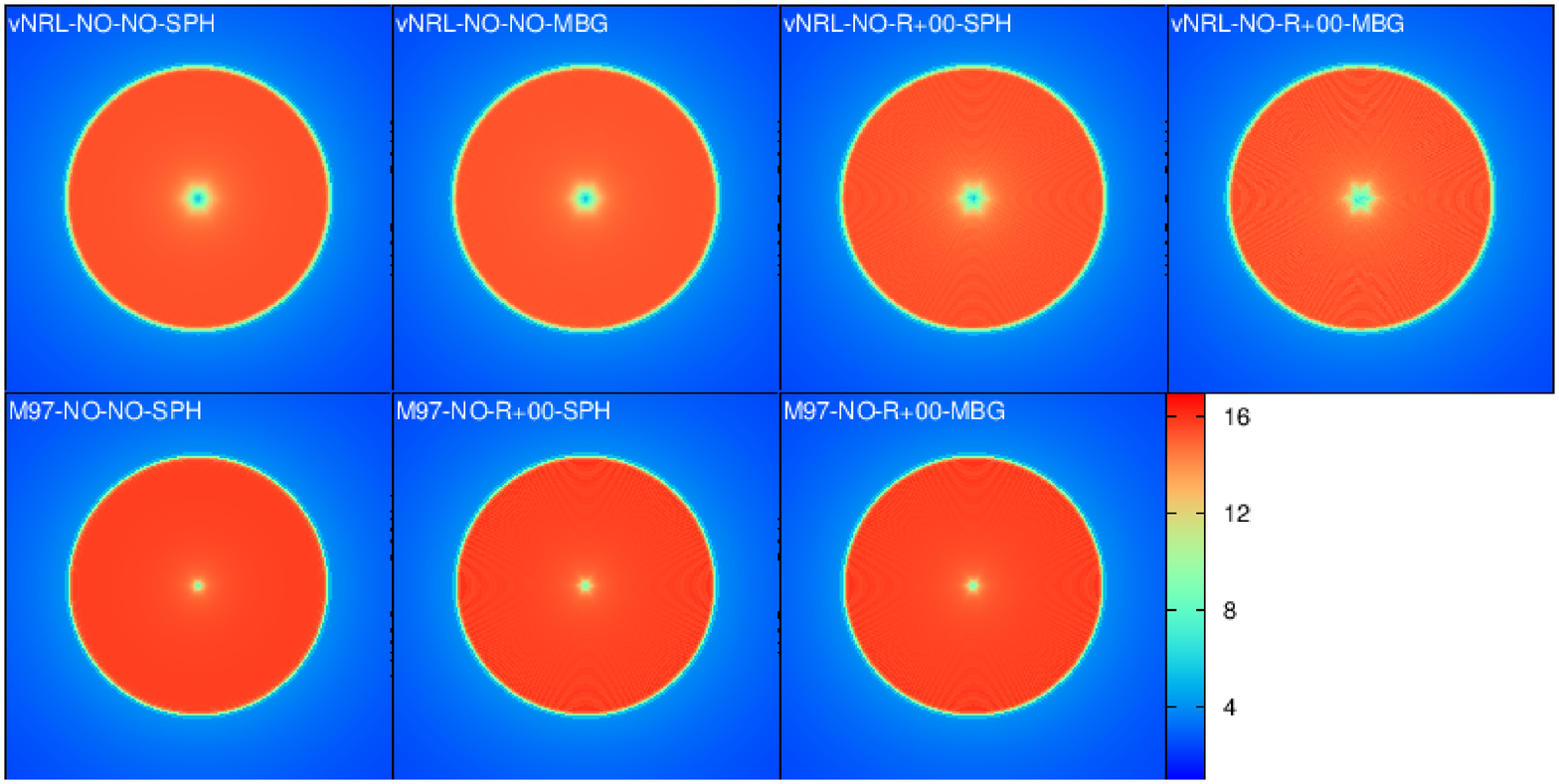}
\caption{
	Snapshots from the Noh implosion test at $t = 1.0$ with each AV implementation.
	The color code indicates the density.
}
\label{fig:Noh_ss}
\end{figure}

Figures \ref{fig:Noh_2D} and \ref{fig:Noh_ss} show the results of the 2D Noh cylindrical implosion test with each AV implementation.
Each AV implementation shows good agreement to the analytic solution.
Around the central region, high density region with circular shape is formed.
However, in this test, the pair-wise AV gives better results than vNRL AV.
vNRL AV gives somewhat broader shock and larger numerical error at the central region, especially with R+00 switch.
However, note that vNRL AV does not lead to catastrophic results.
The results with vNRL AV is roughly comparable to the results with the other methods \citep[e.g., ][]{G+12}.
We conclude that the pair-wise AV has advantages in dealing with strong shocks.
On the other hand, vNRL form has advantages in the case we need to treat shear flow correctly.

\clearpage

\end{document}